\documentstyle[11pt,aaspp4]{article}

\slugcomment{Submitted to The Astronomical Journal}

\lefthead{Schneider et al.}
\righthead{High-Redshift Quasars}

\begin{document}

\title{High-Redshift Quasars Found in Sloan Digital Sky Survey Commissioning
 Data~VI.  Sloan Digital Sky Survey Spectrograph Observations
\footnote{Based on observations obtained with the Sloan Digital
 Sky Survey and with the Apache Point Observatory 3.5-m telescope, which are
 owned and operated by the Astrophysical Research Consortium}
}
\author{
Scott~F.~Anderson\altaffilmark{\ref{UW}},
Xiaohui~Fan\altaffilmark{\ref{IAS}},
Gordon~T.~Richards\altaffilmark{\ref{PennState}},
Donald~P.~Schneider\altaffilmark{\ref{PennState}},
Michael~A.~Strauss\altaffilmark{\ref{Princeton}},
Daniel~E.~Vanden~Berk\altaffilmark{\ref{FNAL}},
James~E.~Gunn\altaffilmark{\ref{Princeton}},
Gillian~R.~Knapp\altaffilmark{\ref{Princeton}},
David~Schlegel\altaffilmark{\ref{Princeton}},
Wolfgang~Voges\altaffilmark{\ref{MPE}},
Brian~Yanny\altaffilmark{\ref{FNAL}},
Neta~A.~Bahcall\altaffilmark{\ref{Princeton}},
J.~Brinkmann\altaffilmark{\ref{APO}},
Robert~Brunner\altaffilmark{\ref{Caltech}},
Istvan~Csab\'ai\altaffilmark{\ref{JHU}}$^,$\altaffilmark{\ref{Hungary}},
Mamoru~Doi\altaffilmark{\ref{Tokyo2}},
Masataka~Fukugita\altaffilmark{\ref{CosJapan}}$^,$\altaffilmark{\ref{IAS}},
\v Zeljko~Ivezi\'c\altaffilmark{\ref{Princeton}},
Donald~Q.~Lamb\altaffilmark{\ref{Chicago}},
Jon~Loveday\altaffilmark{\ref{Sussex}},
Robert~H.~Lupton\altaffilmark{\ref{Princeton}},
Timothy~A.~McKay\altaffilmark{\ref{Michigan}},
Jeffrey~A.~Munn\altaffilmark{\ref{USNOAZ}},
R.C.~Nichol\altaffilmark{\ref{CMU}},
G.P.~Szokoly\altaffilmark{\ref{Potsdam}},
and
Donald~G.~York\altaffilmark{\ref{Chicago}}
}

email addresses: anderson@astro.washington.edu, fan@sns.ias.edu,
gtr@astro.psu.edu, dps@astro.psu.edu, strauss@astro.princeton.edu,
danvb@fnal.gov

\newcounter{address}
\setcounter{address}{2}
\altaffiltext{\theaddress}{University of Washington, Department of
   Astronomy, Box 351580, Seattle, WA 98195.
\label{UW}}
\addtocounter{address}{1}
\altaffiltext{\theaddress}{The Institute for Advanced Study, Princeton,
   NJ 08540.
\label{IAS}}
\addtocounter{address}{1}
\altaffiltext{\theaddress}{Department of Astronomy and Astrophysics, The
   Pennsylvania State University, University Park, PA 16802.
\label{PennState}}
\addtocounter{address}{1}
\altaffiltext{\theaddress}{Princeton University Observatory, Princeton,
   NJ 08544.
\label{Princeton}}
\addtocounter{address}{1}
\altaffiltext{\theaddress}{Fermi National Accelerator Laboratory, P.O. Box 500,
   Batavia, IL 60510.
\label{FNAL}}
\addtocounter{address}{1}
\altaffiltext{\theaddress}{Max-Planck-Institue f\"ur extraterrestrische Physik,
   Postfach~1603, 85750 Garching, Germany.
\label{MPE}}
\addtocounter{address}{1}
\altaffiltext{\theaddress}{Apache Point Observatory, P.O. Box 59,
   Sunspot, NM 88349-0059.
\label{APO}}
\addtocounter{address}{1}
\altaffiltext{\theaddress}{Astronomy Department, California Institute of
   Technology, Pasadena, CA 91125.
\label{Caltech}}
\addtocounter{address}{1}
\altaffiltext{\theaddress}{Department of Physics and Astronomy,
   Johns Hopkins University, 3701 University Drive, Baltimore, MD 21218.
\label{JHU}}
\addtocounter{address}{1}
\altaffiltext{\theaddress}{Department of Physics of Complex Systems,
   E\"otv\"os University, P\'azm\'ay P\'eter \hbox{s\'et\'any 1/A,}
   H-1117, Budapest, Hungary.
\label{Hungary}}
\addtocounter{address}{1}
\altaffiltext{\theaddress}{Department of Astronomy and Research Center for the
   Early Universe, School of Science, University of Tokyo, Mitaka,
   Tokyo 181-0015, Japan.
\label{Tokyo2}}
\addtocounter{address}{1}
\altaffiltext{\theaddress}{Institute for Cosmic Ray Research, University
   of Tokyo, Midori, Tanashi, Tokyo 188-8588, Japan
\label{CosJapan}}
\addtocounter{address}{1}
\altaffiltext{\theaddress}{Astronomy and Astrophysics Center, University of
   Chicago, 5640 South Ellis Avenue, Chicago, IL 60637.
\label{Chicago}}
\addtocounter{address}{1}
\altaffiltext{\theaddress}{Astronomy Centre, University of Sussex, Falmer,
   Brighton BN1 9QJ, UK.
\label{Sussex}}
\addtocounter{address}{1}
\altaffiltext{\theaddress}{Department of Physics,University of Michigan,
   500 East University, Ann Arbor, MI 48109.
\label{Michigan}}
\addtocounter{address}{1}
\altaffiltext{\theaddress}{US Naval Observatory, Flagstaff Station,
   P.O. Box 1149, Flagstaff, AZ 86002-1149.
\label{USNOAZ}}
\addtocounter{address}{1}
\altaffiltext{\theaddress}{Dept. of Physics, Carnegie Mellon University,
     5000~Forbes Ave., Pittsburgh, PA~15232.
\label{CMU}}
\addtocounter{address}{1}
\altaffiltext{\theaddress}{Astrophysikalisches Institut Potsdam, Germany.
\label{Potsdam}}

\vbox{
\begin{abstract}
We present results on over 100 high-redshift quasars found in the
Sloan Digital Sky Survey (SDSS), using automated selection algorithms
applied to SDSS imaging data and with spectroscopic confirmation
obtained during routine spectroscopic operations of
the Sloan~2.5-m telescope.  The SDSS
spectra cover the
wavelength range \hbox{3900 -- 9200 \AA } \ at a spectral resolution of~1800,
and
have been obtained for~116 quasars with redshifts greater than~3.94;~92
of these
objects were previously uncataloged, significantly increasing
the current tally of
published $z>4$ quasars.
The paper also reports observations of
five additional new \hbox{$z > 4.6$} quasars; all were
found from the SDSS
imaging survey and spectroscopically confirmed with data from the Apache
Point Observatory's 3.5-m telescope.
The~$i'$ magnitudes
of the quasars range from \hbox{18.03 to 20.56.}  Of the~97 new objects
in this paper,~13 are Broad Absorption Line quasars.  Five quasars, including
one object at a redshift of~5.11, have
20~cm peak flux densities greater than~1~mJy.  Two of the quasars, both
\hbox{at $z \approx 4.5$}, have very weak emission lines; one of these objects
is a radio source.
Nineteen of the newly-discovered objects 
have redshifts above~4.6, and the maximum redshift is~$z$=5.41; among objects
reported to date, the latter is the third highest redshift AGN, and 
penultimate in redshift among luminous quasars.

\end{abstract}
}



%

\section{Introduction}

The past few years have seen a dramatic increase in both the number of known
high-redshift quasars and in the highest quasar redshift.  Large area
surveys, using multicolor selection techniques, have identified a number
of quasars at redshifts larger than 5.0, including one object
at a redshift of~5.80 (Fan et al.~2000b).  Surveys using photographic
plates (e.g., Kennefick, Djorgovski, \& de~Carvalho~1995b;
Storrie-Lombardi et al.~2000, Sharp et al.~2001)
and CCDs (the Sloan Digital Sky Survey (SDSS); \hbox{York et al.~2000)}
have now produced a data base of well over 100 quasars with redshifts
larger than four.  
Given the current pace of
discovery, we expect the
number of such quasars will increase by many factors in the near future.

In this series of papers we have already presented SDSS
discoveries of more than 100 quasars with redshifts larger than~3.5 (four
quasars with redshifts larger than~4.95); all were initially identified
in SDSS imaging data and spectroscopic confirmation was obtained
using the Apache Point Observatory 3.5-m, Hobby-Eberly, and Keck telescopes
(see Schneider et al.~2001 for a summary of the SDSS high-redshift quasars).
The SDSS spectrographs began operation in early~2000
(see Castander et al.~2001), and in the past year have returned spectra of
nearly~9000 quasars; for the initial results of the SDSS quasar survey
see
Richards et al.~(2001b) and Vanden~Berk et al.~(2001).
In this paper we report the first results of the SDSS spectroscopic
survey for high-redshift quasars: 116 quasars (92 previously unknown)
with redshifts larger
than~3.94 identified by commissioning versions of
the SDSS Quasar Target Selection Software; the final version of this
code is presented in
Richards et al.~(2001a).
In addition, we also describe five
new \hbox{$z > 4.6$} quasars that were found in the SDSS imaging data and
spectroscopically confirmed with the Apache Point Observatory 3.5-m telescope.
Finding charts for all objects lacking published identifications
are given in Figure~1.

The SDSS imaging observations and target selection are described in \S 2,
and the spectroscopic observations of the quasar candidates are
presented in \S 3.
The properties of
the quasars are reviewed in \S 4, and a brief discussion appears in \S 5.
Throughout this paper we will adopt the cosmological model with
\hbox{$H_0$ = 50 km s$^{-1}$ Mpc$^{-1}$,} \hbox{$\Omega_0$ = 1.0,}
\hbox{and $\Lambda$ = 0.0.}

\section{Sloan Digital Sky Survey Imaging and Quasar Target Selection}

The Sloan Digital Sky Survey
uses a CCD camera \hbox{(Gunn et al. 1998)} on a
dedicated 2.5-m telescope \hbox{(Siegmund et al. 2001)}
at Apache Point Observatory,
New Mexico, to obtain images in five broad optical bands over
10,000~deg$^2$ of the high Galactic latitude sky centered approximately
on the North Galactic Pole.  The five filters (designated $u'$, $g'$,
$r'$, $i'$, and~$z'$) cover the entire wavelength range of the CCD
response \hbox{(Fukugita et al. 1996;} \hbox{Fan et al.~2001).}
Photometric calibration is provided by simultaneous
observations with a 20-inch telescope at the same site.  The
survey data processing software measures the properties of each detected object
in the imaging data, and determines and applies both
astrometric and photometric
calibrations (\hbox{Pier et al. 2001}; \hbox{Lupton et al. 2001}).
At the time of this writing (March~2001) substantially more
than~1000~sq.~deg.~have been observed with the SDSS, although some of the data
do not meet the strict survey
requirements.  The finding charts in Figure~1 were made from $i'$-band
data taken with the SDSS survey camera.

The high photometric accuracy, good image quality, and five wavelength bands
covering the optical and near-infrared of the SDSS imaging survey produce
an extremely effective data base from which to identify quasars.  Since the
spectra of quasars and stars differ considerably, multicolor surveys have
long been the primary technique employed to optically select quasars
(e.g., Schmidt \& Green~1983 and references therein).
Fan~(1999) calculated the expected location, as a function of redshift,
of quasars in SDSS color-space; his results suggested
that the SDSS could effectively
identify quasars at most redshifts \hbox{below $\approx$ 5-6.}
Richards et al.~(2001b) presented
the SDSS colors of more than~2600 quasars with \hbox{$0 < z < 5$;}
the data show tight color-redshift relations for the vast
majority of quasars, and the median relation to a large extent
follows the predictions of Fan~(1999).

The presence of the $i'$ and $z'$ filters in the SDSS camera make it possible
to effectively identify high-redshift quasars.
The combination of the strong
Lyman~$\alpha$ emission line, substantial absorption due to the Lyman~$\alpha$
forest, and the smooth continuum longward of the Lyman~$\alpha$ emission line
cause the SDSS colors of most quasars with redshifts above~$\approx$~3.3 to
be considerably different than SDSS colors of stars
(e.g., Fan~1999; Fan et al.~1999; Richards et al.~2001b).  
As one increases the quasar redshift above~$\approx$~5.8, the Lyman~$\alpha$
emission line
leaves the $i'$ band and enters the
$z'$ filter, so detections \hbox{of most $z \approx 6$} quasars will rely
on only one filter (see Fan~2000b).
To date, the SDSS high-redshift quasar selection efficiency
(number of quasars divided by the number of quasar candidates) is
approximately \hbox{60--70\%}
\hbox{($e.g.,$ Fan et al.~2001; Schneider et al.~2001)}
for objects with \hbox{$i^* < 20.0$;} this is a much
higher value than that achieved in previous investigations in this field.
The contaminants in the SDSS high-redshift quasar candidates are
late-type stars, narrow emission line galaxies at low redshift,
and \hbox{``E+A galaxies"} \hbox{at $z \approx 0.4$} (Fan et al.~1999).

One of the primary goals of the SDSS is to obtain spectra of a sample
\hbox{of $\approx$ 100,000 quasars} selected from the SDSS imaging data
(York et al.~2000).  The 
SDSS Quasar Target Selection Algorithm
uses information from the SDSS photometric catalogs (e.g., morphology,
magnitude) to produce a list of quasar candidates to be included in
the SDSS spectroscopic survey.  The details of the selection are given
in Richards et al.~(2001a).  One of the primary tasks of the SDSS
commissioning period was to refine the quasar selection technique;
during this time a number of versions of the quasar selection algorithm
were used.

It is important to note that the objects described in this paper
do not constitute
a complete sample.  Although most of the objects were identified by an
automated selection technique, 
the details of the selection procedure varied from field to field as the
effectiveness of the algorithm was tested during the commissioning period.

The planned SDSS complete sample is expected to consist
of objects whose colors are distinct from those of stars with
\hbox{$15 < i' < 19.2$} (the bright limit is set to avoid  saturation properties
in the SDSS spectrographs), objects with the colors of \hbox{$z > 3.0$}
quasars with \hbox{$15 < i' < 20.5$,}
and point sources with \hbox{$i' < 19.2$}
that are coincident with radio sources in the FIRST survey (Becker, White, \&
Helfand~1995).  Low-redshift quasar candidates can have a nonstellar
appearance.  The expectation is that the quasar survey will have
an efficiency (quasars:quasar candidates) of at least~65\% and a completeness
of at least~90\% (i.e., 90\% of previously known quasars will be recovered).

\section{Spectroscopy of Quasar Candidates} \label{survfields}

The SDSS spectroscopic survey is carried out by two fiber-fed double
spectrographs mounted at the Cassegrain focus of the SDSS 2.5-m telescope
(see York et al.~2000; Castander et al.~2001; and
Uomoto et al.~2001 for details).
Each spectrograph contains blue \hbox{(3900-6200 \AA )} and red
\hbox{(5800-9200 \AA )} beams that produce spectra at a resolution
\hbox{of $\approx$ 1800.}  A total of 320 fibers enter each spectrograph;
a single observation covers targets located in a~3$^{\circ}$ diameter field.
The fibers subtend a diameter of~3$''$ on the sky, and because of mechanical
constraints the fibers must be separated by at least~55$''$.  In each field,
32 of the 640 fibers
are assigned to measuring the sky, and approximately eight fibers are used for
photometric standards and three fibers observe reddening standards.
Typically about~100 quasar
candidate spectra are taken in a 45-minute observation of a field.
For this paper we considered spectroscopic observations of 138 fields,
which cover an effective area of approximately~700~sq~deg; nearly 9000
quasar spectra have been obtained to date.

The data, along with the associated calibration frames, are processed by
the SDSS Spectroscopic Pipeline (Frieman et al.~2001), which removes
instrumental effects, extracts the spectra, calculates the wavelength
calibration, subtracts the sky spectrum and removes the atmospheric
absorption bands, and performs the flux calibration.  The spectra are 
then classified (e.g., star, galaxy, quasar) and redshifts are determined
by the pipeline software.

We selected objects
whose redshifts, as determined by the SDSS spectroscopic pipeline software,
were larger than~3.95, and supplemented this list by
visually inspecting the processed spectra and identifying objects whose
redshifts appeared to be four or larger.
A visual inspection of these candidates produced a set of
116 quasars with redshifts
ranging from~3.94 to~5.41.  Although the C~IV emission line lies beyond
the red cutoff of the spectrograph when the redshift exceeds~4.8, the
quality of the spectra and the distinctive appearance of the
\hbox{Lyman~$\alpha$+N V} emission line and Lyman~$\alpha$ forest region
in these objects allows single-line redshifts to be assigned with
high confidence.  SDSS spectra of~27 of the
quasars
are displayed in Figure~2.

The high quality of the data is particularly notable given that the spectra,
of objects with \hbox{$i^* \approx 20$,} were obtained with a 2.5-m
telescope in modest length exposures (less than an hour
in clear, good seeing conditions).

An additional five \hbox{$z > 4.6$} quasars, identified as high-redshift
candidates in the SDSS imaging data, were
spectroscopically confirmed with
the Double Imaging Spectrograph on the Apache Point Observatory 3.5-m telescope.
The spectra covered the wavelength range 4000-10,000~\AA , with a spectral
resolution of 12~\AA\ (blue camera) and 23~\AA\ (red camera); all
exposure times were 3600~s.  (For details of the instrument configuration
see Fan et al.~1999).  Redshifts for these five objects were determined
using the techniques described in Fan et al.~(2001).  The spectra
of the five quasars are shown in Figure~3.

\section{Discussion}

Table~1 provides basic data (uniform and of high quality) for all 121 quasars.
The object name format is
\hbox{SDSSp Jhhmmss.ss+ddmmss.s}, where the coordinate
equinox is~J2000, and the ``p" refers to the preliminary nature of the
astrometry.  The reported magnitudes
are based on a preliminary photometric calibration; to indicate this,
the filters have an asterisk instead of a prime superscript (e.g., $g^*$
rather than~$g'$).  The
estimated astrometric accuracies in each coordinate are~0.10$''$ and the
calibration of the photometric measurements is accurate to~0.04~magnitudes
in the~$g'$, $r'$, and~$i'$ filters
and~0.06~magnitudes in the $u'$ and~$z'$ bands.
Throughout the text, object names will frequently
be abbreviated \hbox{as SDSShhmm+ddmm.}

Slightly more than~80\% (97)
of the quasars in Table~1 were previously unknown, including
all five of the $z > 4.6$ ``APO" objects.  Eighteen
of the objects had been previously identified in SDSS data; they are included
in this study because of the improved photometry and the newly obtained
SDSS spectra.  Four quasars in Table~1 were discovered by the Automated
Plate Measuring facility (APM) survey (see Storrie-Lombardi et al.~2000),
and two quasars are Palomar Sky Survey (PSS) objects 
(Kennefick, Djorgovski, \& de~Carvalho~1995b).
The final column in Table~1 provides
the references for all~24 of the previously known quasars independently
recovered here.

The locations of the quasars in SDSS color space are shown in Figure~4.
As expected from the predictions of Fan~(1999) and demonstrated by previous
SDSS high-redshift quasar studies (e.g., see summary in Schneider et al.~2001),
the colors of quasars with redshifts between 4.0 and~4.6 are
well-separated from the colors of stars in
the \hbox{$(g^* - r^*),(r^* - i^*)$} diagram (the distance between
stars and a typical quasar at this redshift is approximately one magnitude).
At redshifts above~$\approx$~4.6 the \hbox{$(r^* - i^*),(i^* - z^*)$}
diagram becomes a very useful aid, because at these redshifts the errors
in the $g^*-r^*$ color can become large due to Lyman-limit systems
entering the~$g^*$ band.

Figure~5 shows the relationship between the $(r^*-i^*)$ color and redshift
for the~121 quasars in Table~1.  Between redshifts of~4.0 and~4.5, the
mean color slightly increases, and the colors in this redshift range
have a dispersion about
the mean
\hbox{of 0.10-0.15 mag.}  At redshifts above~4.5,
when the Lyman~$\alpha$ emission line
moves from the~$r'$ to the~$i'$ filter and the Lyman~$\alpha$ forest
blankets the~$r'$ bandpass, the $(*r^*-i^*)$ measurements rapidly
increase to~1.5 and larger.  The large dispersion in the color at~$z>4.5$
is likely to arise from the variation in strength of the Lyman~$\alpha$ forest
along different lines of sight.

The redshift distribution of the quasars in Table~1 is shown in Figure~6.
The numbers decline rapidly with redshift (only about~10\% of the objects
have \hbox{$z > 4.8$)}, and there is a slight dip near \hbox{$z = 4.5$}
due to the difficulty of separating quasars of this redshift from the
stellar locus (see Fan et al.~2001) using the early commissioning selection
algorithm.  Figure~7 displays the distribution
of~$i^*$~magnitudes of the quasars.  The bulk of the objects
have \hbox{$19.2 < i^* < 20.4$}; the brightest object has \hbox{$i^* = 18.03$}
and the faintest quasar has \hbox{$i^* = 20.56$.}  Fourteen of the quasars
\hbox{have $i^* < 19.0$.}

These objects are moderately luminous quasars; 3C~273, which has
\hbox{$M_B = -27.0$} in our adopted cosmology, would appear in our sample
\hbox{($i^* < 20.6$)} out to a redshift of approximately~4.8.  The most
luminous objects in Table~1 are over two magnitudes brighter than~3C~273.

The sample includes nineteen new quasars with $z>4.6$, including
three new quasars with $z>5$: SDSS0231$-$0728, SDSS0756$+$4104, and 
SDSS0913$+$5919. 

The sample contains one set of quasars that are relatively close together
on the sky:  SDSS1108$-$0059 ($z = 4.01$)
and SDSS1108$-$0058 ($z = 4.56$) are separated by only~80$''$.  The projected
comoving separation of the lines of sight \hbox{at $z = 4.01$}
is approximately~500~kpc.

Sixteen of the 121 quasars, and 13 of the 97 newly discovered ones,
are either certain or probable BAL quasars; this is consistent with
the~10\% BAL fraction
found in optically selected quasars at lower redshift
(Weymann et al.~1991).  The spectra of many of the quasars have narrow
absorption line systems, both intervening and intrinsic.

Five quasars in Table~1 have peak 20~cm (FIRST) flux densities
greater than~2.5~mJy.  Two objects which are not included in the FIRST
survey area were detected by the NRAO VLA Sky Survey
(NVSS, Condon et al.~1998; see notes on individual objects).  An additional
two quasars were detected by FIRST, but were fainter than than the NVSS
limit.
Four of the radio quasars were previously known; only five of the 97 newly
discovered quasars are radio sources above the~$\approx$~1~mJy level.
One of five quasars, SDSS0913+5919 \hbox{($z$ = 5.11),} is
a luminous radio source (18.1~mJy at 20~cm), and is the most distant
known radio-loud quasar.
This fraction of radio detections \hbox{$(\le$ 10\%)}
is comparable to the results found
in previous studies of high-redshift quasars (e.g., Schmidt et al.~1995b,
Stern et al.~2000a).

None of the quasars are found in the ROSAT All-Sky Survey (RASS)
Bright Source Catalog
(Voges et al.~1999; Schwope et al.~2000), which
has a flux limit of \hbox{$2.4 \times 10^{-12}$ erg cm$^{-2}$ s$^{-1}$}
in the~0.5 to~2.0~keV band.  This result is not unexpected given that with
the exception of a few unusual sources (e.g., blazars), high-redshift quasars
have X-ray fluxes well below the catalog limit
(Kaspi, Brandt, \& Schneider~2000).  One quasar, SDSS1737+5828, may have been
detected in the RASS Faint Source Catalog 
(Voges et al.~2000), but the optical/X-ray match is far from certain
(see notes on individual objects).

\bigskip
\centerline{Notes on Individual Objects}

\medskip

\medskip

\vbox{\noindent
{\bf SDSSp J003126.80+150739.6} ($z = 4.20$): This radio-loud quasar
has a flux density of~42.2~mJy at 20~cm from the NVSS.
}

\medskip

\vbox{\noindent
{\bf SDSSp J012004.83+141108.3} ($z = 4.71$): This object contains deep
C~IV and Si~IV BAL features, and there may be strong \hbox{Lyman~$\beta$+O VI}
emission.  The spectrum was taken with the APO~3.5-m telescope; the data
for this object are displayed in Figure~3.
}

\medskip

\vbox{\noindent
{\bf SDSSp J015032.87+143425.6} ($z = 4.14$): The spectrum of this quasar shows
very deep, very broad Si~IV and C~IV absorption troughs, and also has a
narrow, associated absorption line system.
}

\medskip

\vbox{\noindent
{\bf SDSSp J015642.11+141944.4} ($z = 4.30$): Strong BAL features are
apparent in this spectrum, and lines from a given ion have multiple,
broad components.
}

\medskip

\vbox{\noindent
{\bf SDSSp J023137.65$-$072854.5} ($z = 5.41$): This object has
the highest redshift in our current sample, and is the AGN with the third 
largest known redshift (following the luminous SDSS quasar with $z=5.80$
found by Fan et al.~2000b, and a low luminosity AGN at $z=5.50$
found by Stern et al.~2000b).
While the redshift relies especially on the
strong line at \hbox{$\lambda$ = 7800 \AA ,}
the redshift is not in doubt given the characteristic line profile and
continuum discontinuity, plus the presence of a Lyman~limit system
at a rest wavelength of~910~\AA .  There is also strong evidence for
\hbox{Lyman~$\beta$+O VI} and N~V emission at a consistent redshift.
The absorption just shortward of the
Lyman~$\alpha$ emission line is exceptionally strong; the flux appears
to drop to zero.  This object was observed twice with
the SDSS spectrographs, and the two spectra both indicate a redshift
of~5.41.
}

\medskip

\vbox{\noindent
{\bf SDSSp J023923.47$-$080105.1} ($z = 4.00$): This possible
BAL quasar appears to have shallow,
high-velocity C~IV absorption and exceptionally strong
\hbox{O~IV+Lyman~$\beta$} emission.
}

\medskip

\vbox{\noindent
{\bf SDSSp J024447.79$-$081606.1} ($z = 4.03$): \hbox{At $i^* = 18.03$,}
this quasar is both the brightest and the most luminous
\hbox{($M_B \approx -29.3$,} assuming an ultraviolet-optical spectral
index of~$-0.5$) object in the sample.
}

\medskip

\vbox{\noindent
{\bf SDSSp J024457.19$-$010809.9} ($z = 3.96$): This bright
\hbox{($i^* = 18.33$)} BAL quasar is an unpublished PSS object.
}

\medskip

\vbox{\noindent
{\bf SDSSp J025204.29+003136.9} ($z = 4.10$): The spectrum of this
object displays deep, broad C~IV absorption and strong, multi-component
Si~IV absorption.
}

\medskip

\vbox{\noindent
{\bf SDSSp J025647.06$-$085041.4} ($z = 4.21$): This is a BAL quasar
with weak absorption features, the most prominent of which is probably
due to~N~V.
}

\medskip

\vbox{\noindent
{\bf SDSSp J033005.32$-$053709.0} ($z = 4.09$): The spectrum of this
BAL quasar shows two narrow C~IV features superposed on a broad
C~IV absorption trough.
}

\medskip

\vbox{\noindent
{\bf SDSSp J034946.61$-$065730.3} ($z = 3.95$): The spectrum of this
BAL quasar has broad, shallow Si~IV and C~IV absorption features.
}

\medskip

\vbox{\noindent
{\bf SDSSp J075618.4+410408.6} ($z = 5.09$): This quasar has the
fourth largest redshift discovered by the SDSS.
}

\medskip

\vbox{\noindent
{\bf SDSSp J083946.22+511202.8} ($z = 4.39$): 
The FIRST survey measurement of a 20~cm peak flux density of~40.50~mJy
indicates that this is a radio-loud quasar.
}

\medskip

\vbox{\noindent
{\bf SDSSp J085210.89+535949.0} ($z = 4.20$): This is a BAL quasar with
a deep C~IV absorption feature.
}

\medskip

\vbox{\noindent
{\bf SDSSp J085634.93+525206.4} ($z = 4.72$): This object is a
BAL quasar; it has a series of broad, deep absorption features.
The precise redshift is uncertain given the nature of the BAL features.
}

\medskip

\vbox{\noindent
{\bf SDSSp J091316.56+591921.5} ($z = 5.11$): This quasar has the second
largest redshift in our sample, and is near the faint limit
of the SDSS spectroscopic survey.  The spectrum (see Figure~2) is of
modest signal-to-noise ratio, and there is only one significant emission
feature, but the spectral region centered on~$\approx$~7500~\AA\ displays all
the characteristics of Lyman~$\alpha$ emission line at~$z$~=~5.11.  The
radio flux from the object is~18.1~mJy (NVSS); this is a powerful radio
quasar.
}

\medskip

\vbox{\noindent
{\bf SDSSp J102043.83+000105.8} ($z = 4.30$): 
This quasar was detected
by the FIRST survey at a peak flux density of~1.68~mJy at 20~cm.
}

\medskip

\vbox{\noindent
{\bf SDSSp J110826.32+003706.8} ($z = 4.41$): This is a BAL quasar
with strong, smooth Si~IV and C~IV absorption features.
}

\medskip

\vbox{\noindent
{\bf SDSSp J123937.18+674020.8} ($z = 4.40$): The spectrum of this object
displays C~IV absorption and what may be an N~V absorption feature;
we tentatively classify this as a BAL quasar.
}

\medskip

\vbox{\noindent
{\bf SDSSp J130216.13+003032.1} ($z = 4.50$): This object has very
weak emission lines; the spectrum appears quite
similar to the
\hbox{$z = 4.62$} object reported by Fan et al.~(1999).  
}

\medskip

\vbox{\noindent
{\bf SDSSp J141332.36$-$004909.7} ($z = 4.14$): It is not clear whether
this object is a {\it bona fide} BAL or if the C~IV and Si~IV absorption
features are formed from narrow multi-component ``associated" absorption.
The \hbox{Lyman~$\beta$+O~VI} emission line is exceptionally strong.
}

\medskip

\vbox{\noindent
{\bf SDSSp J144231.73+011055.3} ($z = 4.56$): 
This quasar was detected
by the FIRST survey at a peak flux density of~1.07~mJy at 20~cm.  The
spectrum is nearly bereft of emission lines, similar to the
\hbox{$z = 4.62$} object reported by Fan et al.~(1999); also compare
to SDSS1302+0030 noted above.
}

\medskip

\vbox{\noindent
{\bf SDSSp J160501.21$-$011220.6} ($z = 4.92$): This quasar was discovered
by Fan~et al.~(2000a); it is the highest redshift BAL yet published,
and has the by far largest value of $r^* - i^*$ \hbox{(2.37 $\pm$ 0.15)}
in our sample.
}

\medskip

\vbox{\noindent
{\bf SDSSp J162048.74+002005.7} ($z = 4.09$): The spectrum of this quasar
displays a wealth of absorption phenomena: strong, broad C~IV, multi-component
Si~IV, perhaps \hbox{Lyman $\alpha$+N V,} and strong, broad O~IV.
}

\medskip

\vbox{\noindent
{\bf SDSSp J173744.87+582829.5} ($z = 4.94$): This quasar is
possibly X-ray detected
in the RASS Faint Source Catalog (1RXS~J173739.9+582823) with an
unabsorbed flux in the \hbox{0.1--2.4 keV} band
\hbox{of $7.3\times10^{-14}$ erg s$^{-1}$ cm$^{-2}$.} The centroid of
this X-ray source \hbox{($\approx$ 7 photons)} is~40$''$ from the
SDSS position, and the chance of 
at least one unrelated RASS X-ray source superposition in our
sample of over one hundred quasars is non-negligible \hbox{(about 25\%).} 
A higher spatial resolution X-ray image (e.g., from Chandra) would clarify the 
identification. If confirmed, SDSS1737+5828 would
be the highest redshift 
X-ray source yet detected in the RASS.
}

\section{Summary}

Calculations based on the measured high-redshift luminosity function
(e.g., Schmidt, Schneider \& Gunn~1995a; Kennefick, Djorgovski \&
de~Carvalho~1995a) and the expected performance of the SDSS camera
predicted that the SDSS should produce a sample of
well over a thousand quasars
at redshifts above four (e.g., Schneider~1999), and that the SDSS
automated selection algorithm could identify
quasars as distant as a redshift of six.  The early results, presented
in this paper, are promising. Although less than~10\% of the survey area
has been observed spectroscopically, and much of the work to date has
been spent fine-tuning both the equipment and the software, the SDSS has
already, in ``survey mode", selected more than 100 such objects.
We are particularly encouraged by the discovery of the quasar at redshift~5.41;
this object, and the other \hbox{$z \approx 5.0$} quasars in this paper,
show that such objects can be found in an automated manner.

The SDSS commissioning was completed in Fall~2000, and the survey has now
officially commenced.  The quasars in this paper do not constitute
a complete sample because they were identified with a variety of
selection criteria during the commissioning period.
We expect that in each year of operation the SDSS will discover
150-200 quasars will redshifts larger than four, and five to ten quasars
with redshifts larger than five, all found with well-defined, uniform selection
criteria.

\acknowledgments

This work was supported in part by National Science Foundation grants
PHY00-70928~(XF), AST99-00703~(GTR and~DPS), and AST00-71091~(MAS).
XF and MAS acknowledge
additional support from the Princeton University
Research Board, and a Porter O.~Jacobus Fellowship, and XF acknowledges
support from a Frank and Peggy Taplin Fellowship.

The Sloan Digital Sky Survey
\footnote{The SDSS Web site \hbox{is {\tt http://www.sdss.org/}.}}
(SDSS) is a joint project of The University
of Chicago, Fermilab, the Institute for Advanced Study, the Japan
Participation Group, The Johns Hopkins University, the
Max-Planck-Institute for Astronomy (MPIA), the Max-Planck-Institute for
Astrophysics
(MPA), New Mexico State University, Princeton University, the United
States Naval Observatory, and the University of Washington. Apache
Point Observatory, site of the SDSS telescopes, is operated by the
Astrophysical Research Consortium (ARC). 
Funding for the project has been provided by the Alfred~P.~Sloan
Foundation, the SDSS member institutions, the National Aeronautics and
Space
Administration, the National Science Foundation, the U.S.~Department of
Energy, Monbusho, and the Max Planck Society.

\clearpage

%
%

%
%
\newpage
\centerline{\bf Figure Captions}

\figcaption{
$(a$-$d)$ Finding charts for the~98 quasars in this paper that lack published
identifications; each individual
chart is~100$''$ on a side.  All frames are~$i'$ images taken with the SDSS
camera.  The small arrow in the lower left of each chart indicates the
direction of north; all charts have ``sky" parity, so east is
located~90$^{\circ}$ counterclockwise from north.
\label{fig1}}

\figcaption{
$(a$-$i)$ Spectra of~27 of the quasars
taken with the SDSS spectrographs.
The spectral resolution is approximately~1800; data points shortward
of 4250~\AA\ are not displayed as they have very low signal-to-noise ratio.
Displayed are all the \hbox{$z > 4.8$} objects, BALs, and the two
\hbox{$z \approx 4.5$} quasars with extremely weak emission lines.
\label{fig2}}

\figcaption{
Apache Point Observatory 
Spectra of the five $z > 4.6$ quasars taken with the Double Imaging Spectrograph
on the Apache Point Observatory's 3.5-m telescope.  The spectral
resolution is approximately 23~\AA , and the exposure times are all~3600~s.
Only data from the red beam are displayed as these objects produce little
signal below~5500~\AA .
\label{fig3}}

\figcaption{
Locations of the~121 quasars in the
\hbox{$(g^*-r^*),(r^*-i^*)$} (left panel)
and \hbox{$(r^*-i^*),(i^*-z^*)$} (right panel)
color-color diagrams.  The quasars are coded by redshift; triangles
represent redshifts less than~4.6 and circles are $z > 4.6$ quasars.
The points and contours represent the colors of 10,000 stars.
\label{fig4}}

\figcaption{
The $(r^* - i^*)$ color of the 121 quasars as a function of redshift.
Note the sudden reddening at \hbox{$z \approx 4.5$} as the Lyman~$\alpha$
emission line enters the~$i'$ band.
\label{fig5}}

\figcaption{
The redshift distribution of the~121 quasars.  Note the steep decline as one
moves to larger redshifts.  The dip \hbox{at $z \approx 4.5$} is due
to a drop in the detection efficiency at these redshifts
of the quasar selection algorithm used during SDSS commissioning
(see Fan et al.~2001).
\label{fig6}}

\figcaption{
The distribution of the $i^*$ magnitudes of the~121 quasars.  The sharp
cutoff \hbox{near $i* \approx 20.5$} is due to the brightness limit
of the SDSS spectroscopic survey.  Fourteen of the objects
\hbox{have $i^* < 19$.}
\label{fig7}}
\clearpage

\begin{figure}
\plotfiddle{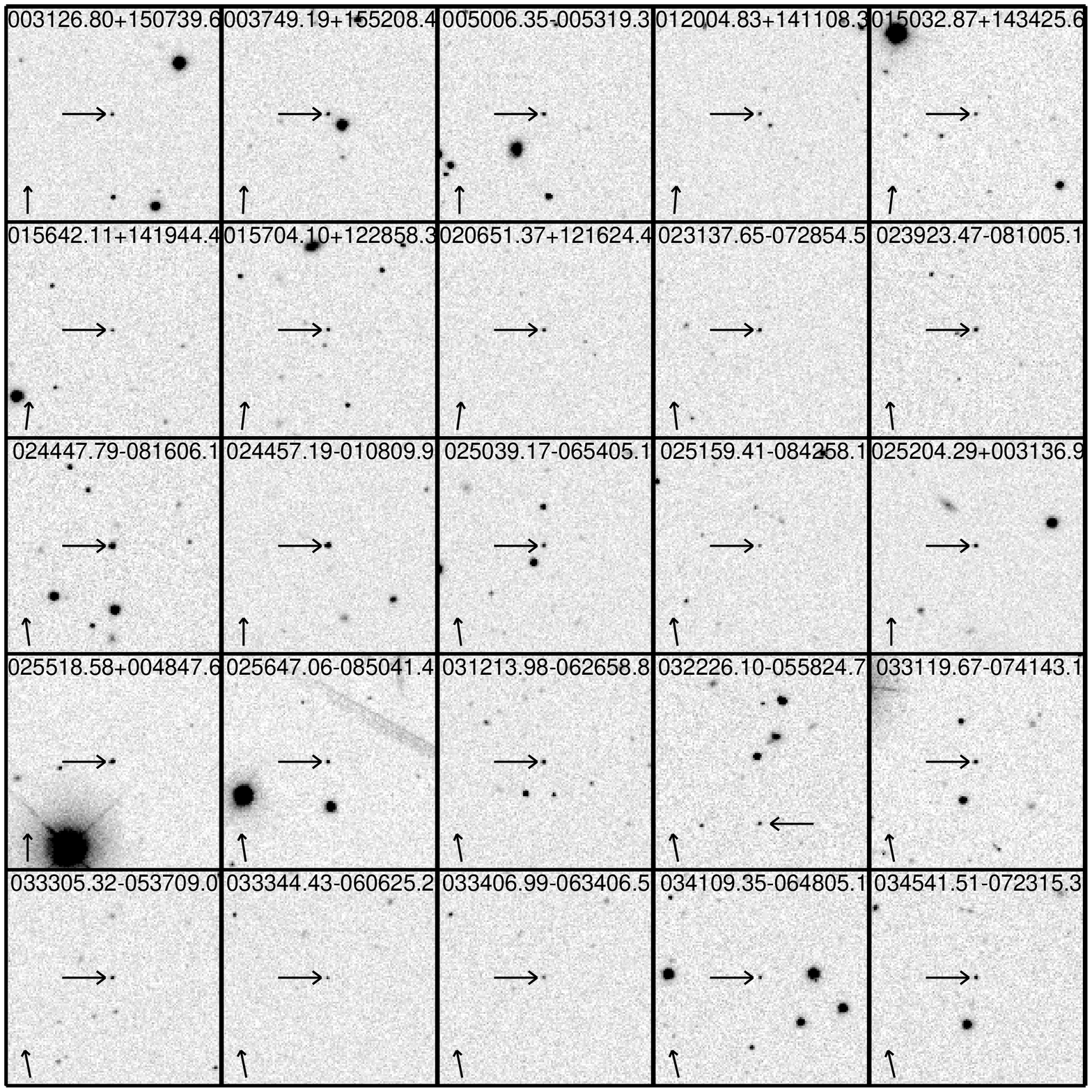}{8.0in}{0.0}{95.0}{95.0}{-290.0}{-40.0}
\end{figure}
\clearpage

\begin{figure}
\plotfiddle{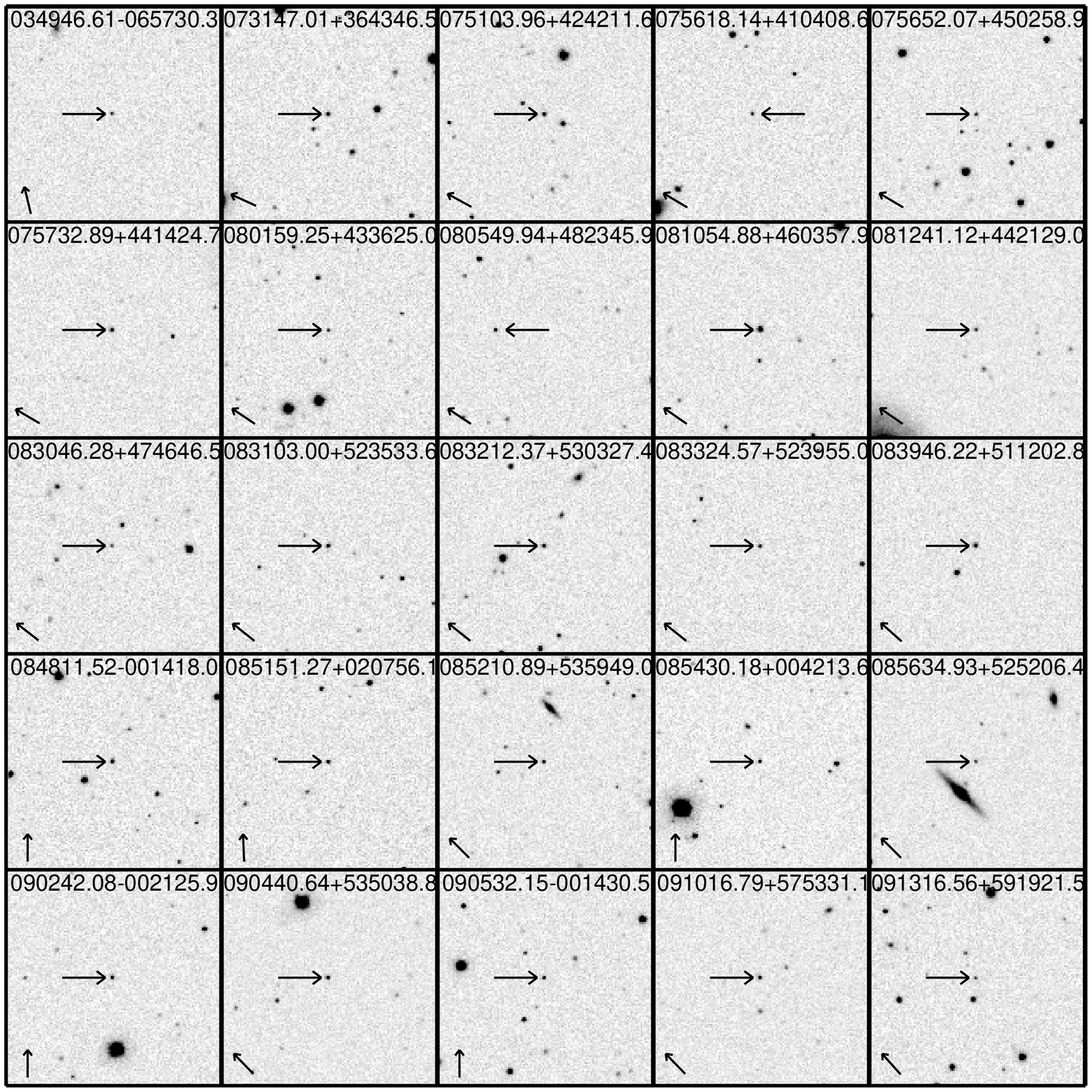}{8.0in}{0.0}{95.0}{95.0}{-290.0}{-40.0}
\end{figure}
\clearpage

\begin{figure}
\plotfiddle{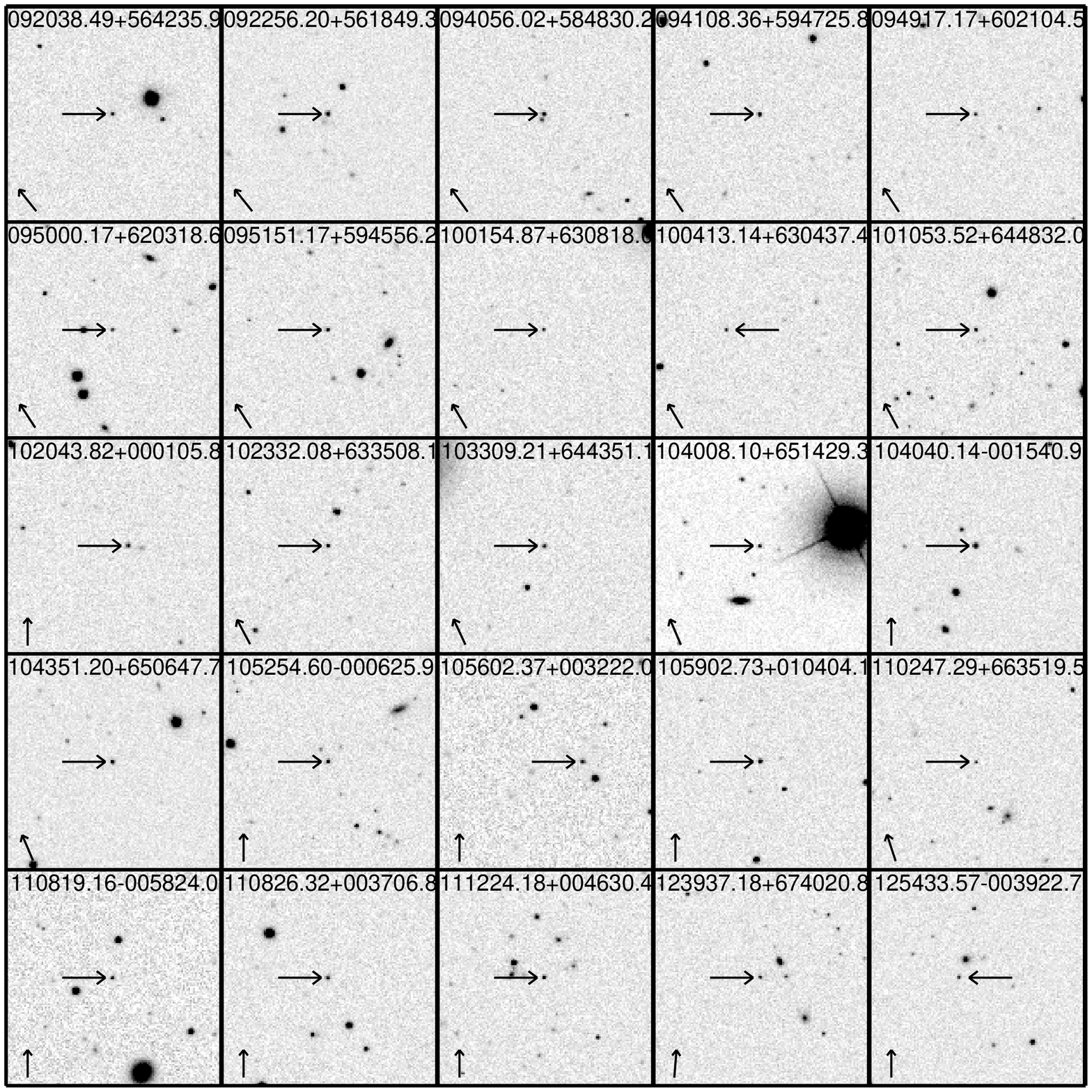}{8.0in}{0.0}{95.0}{95.0}{-290.0}{-40.0}
\end{figure}
\clearpage

\begin{figure}
\plotfiddle{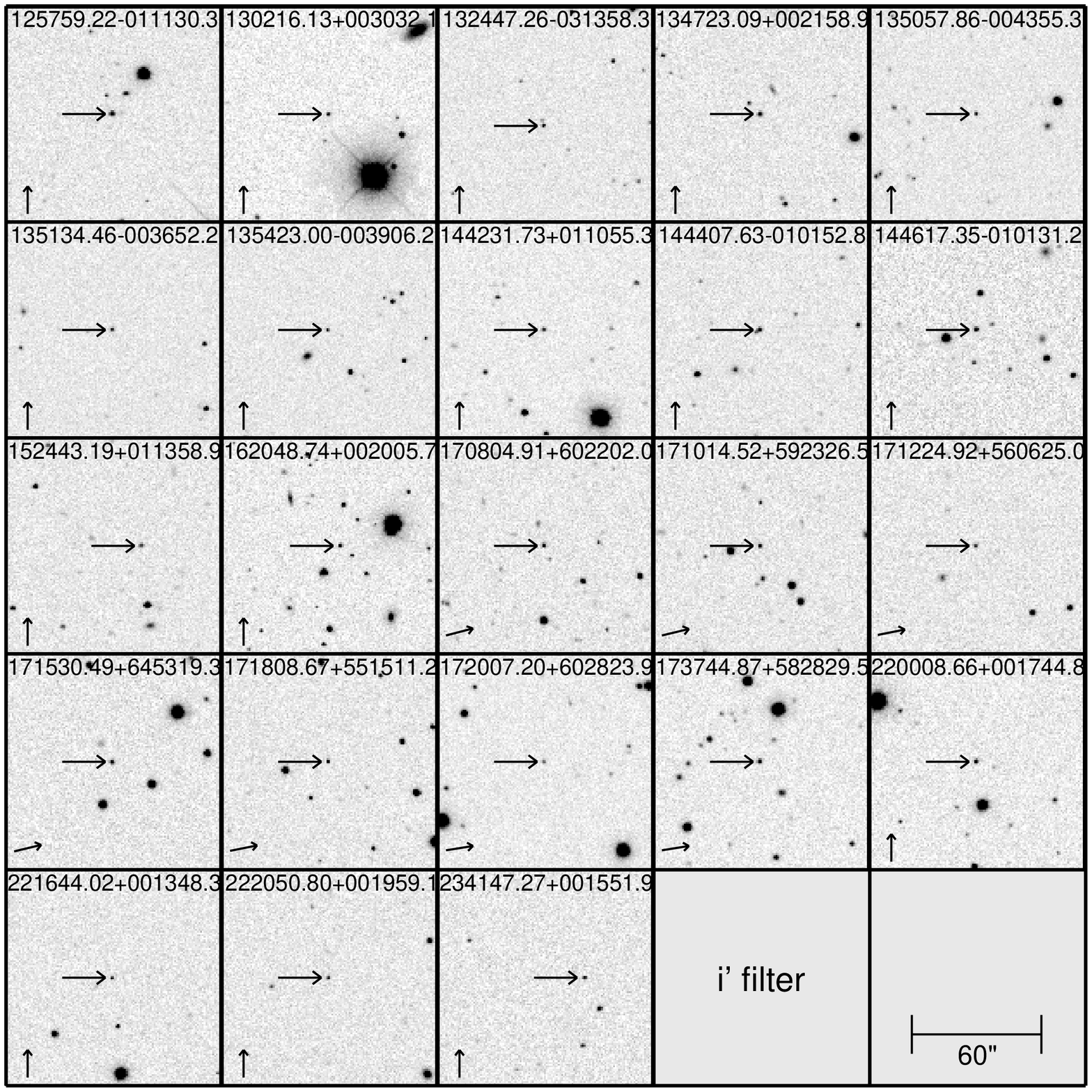}{8.0in}{0.0}{95.0}{95.0}{-290.0}{-40.0}
\end{figure}
\clearpage

\begin{figure}
\plotfiddle{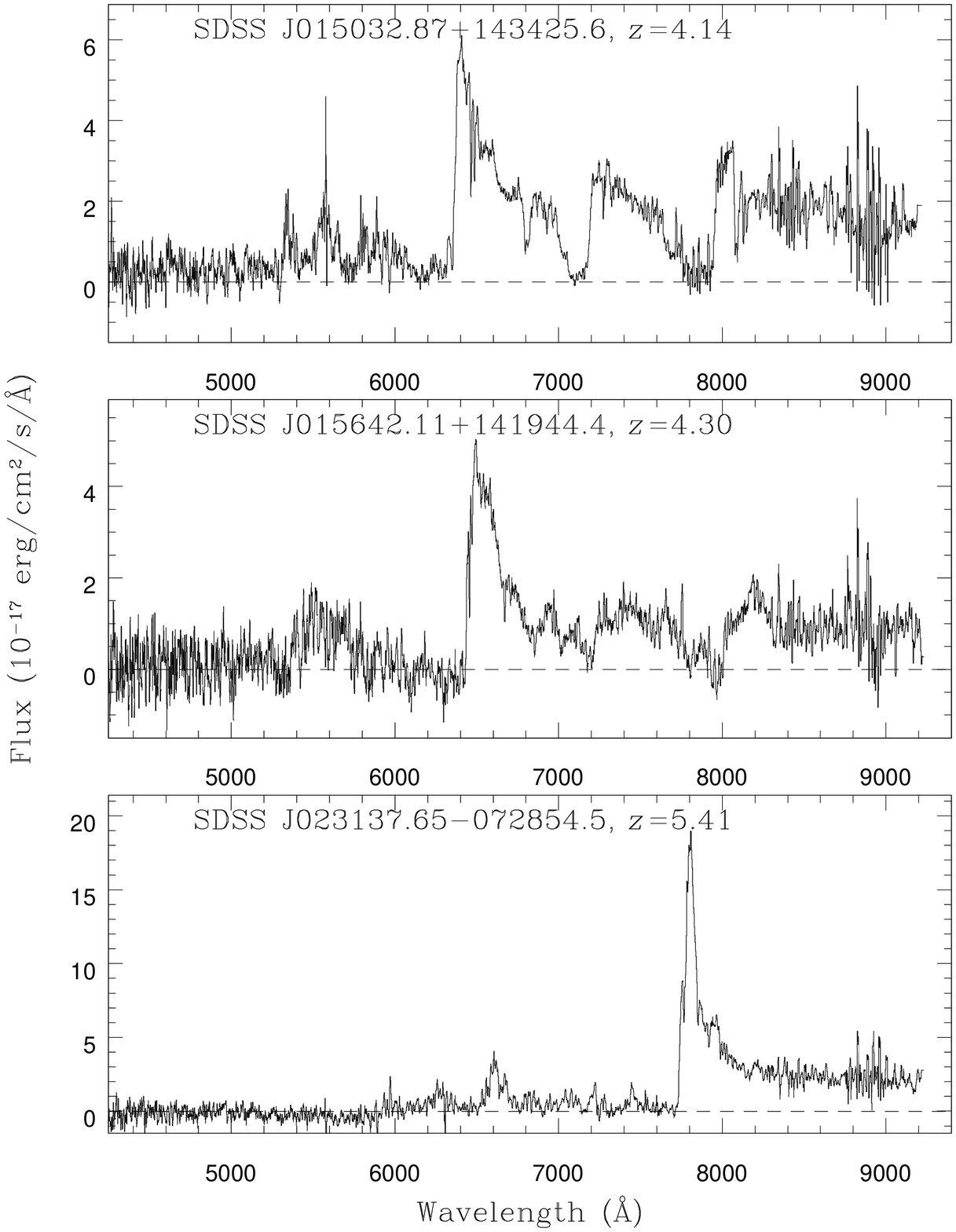}{8.0in}{0.0}{80.0}{80.0}{-250.0}{0.0}
\end{figure}
\clearpage

\begin{figure}
\plotfiddle{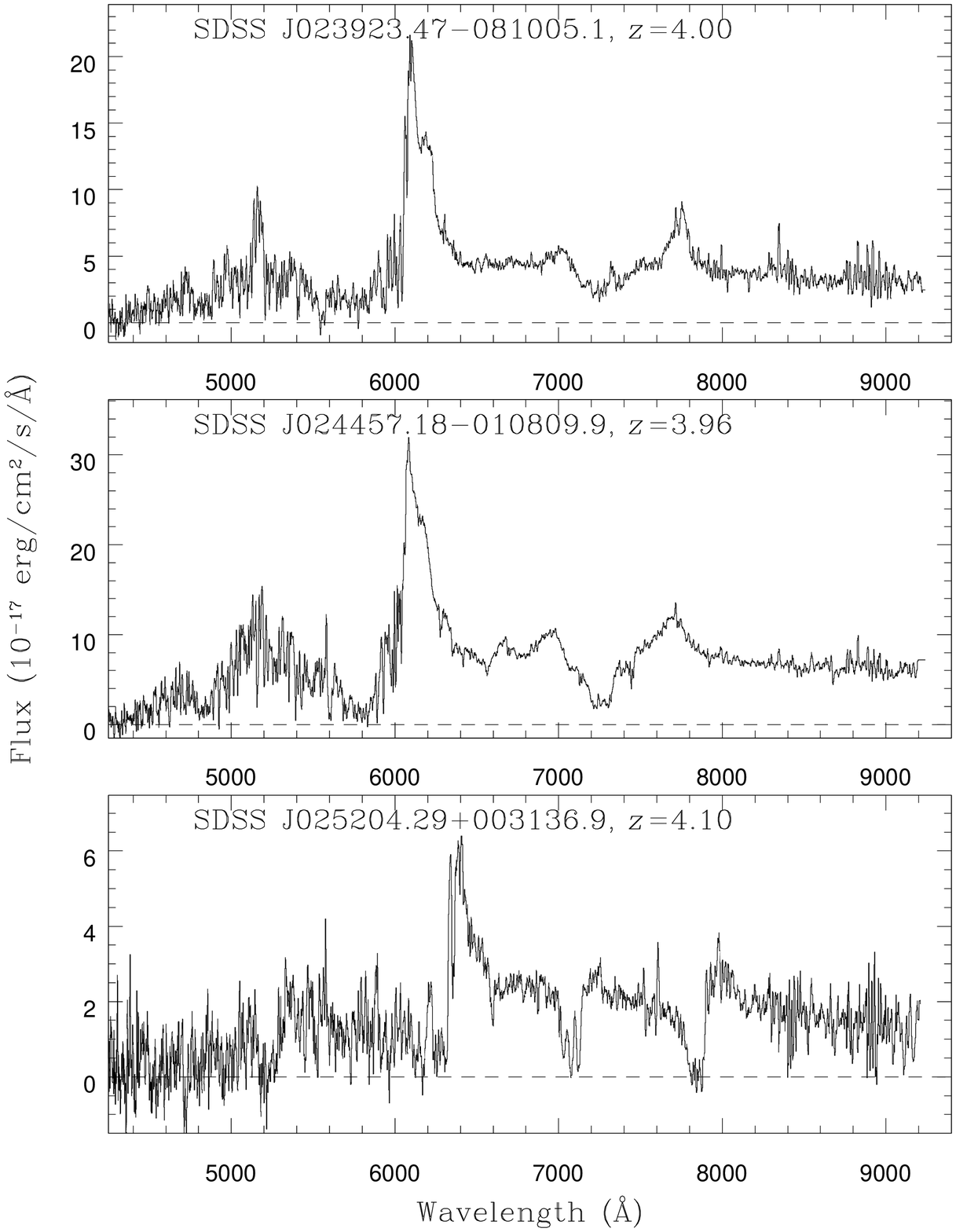}{8.0in}{0.0}{80.0}{80.0}{-250.0}{0.0}
\end{figure}
\clearpage

\begin{figure}
\plotfiddle{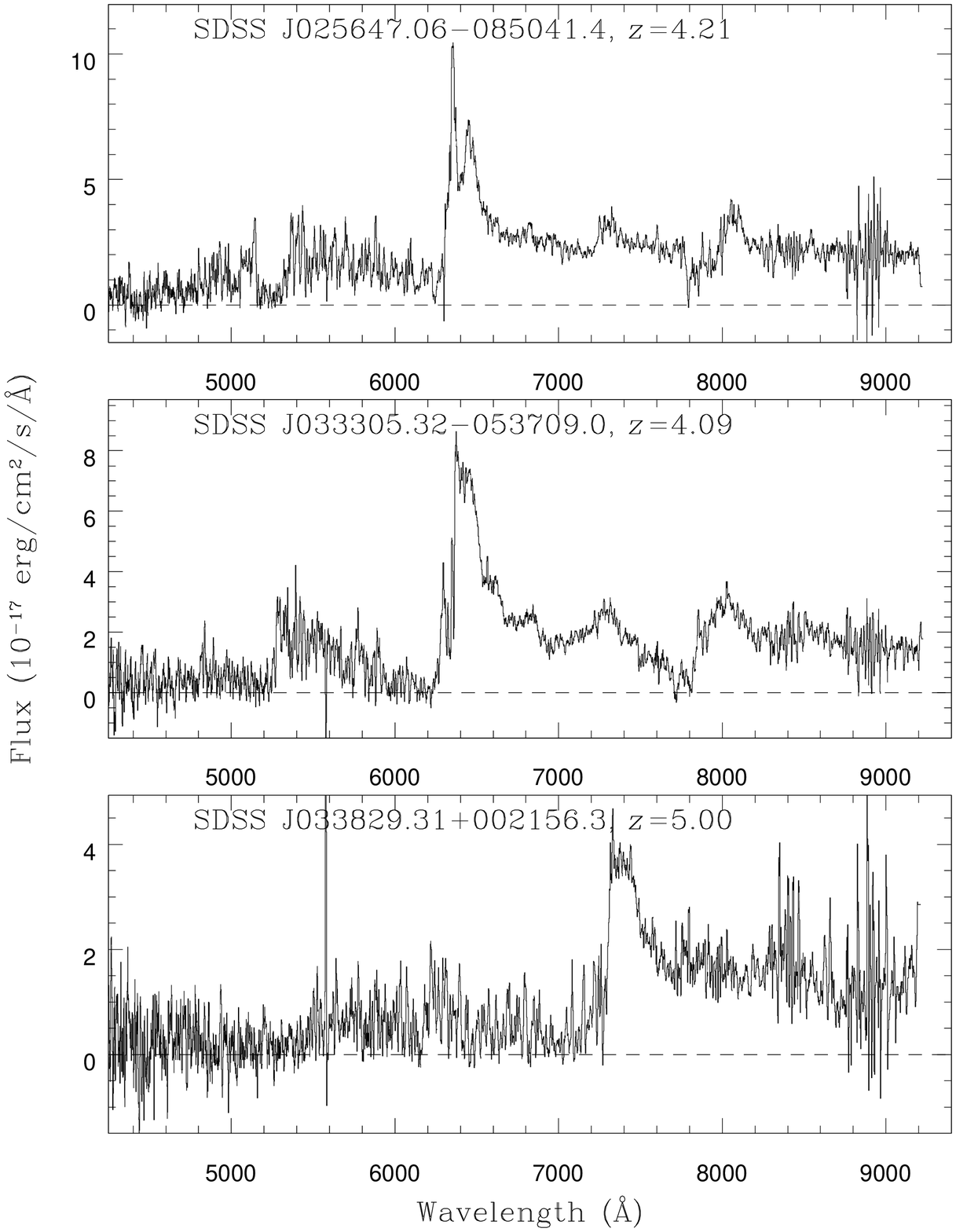}{8.0in}{0.0}{80.0}{80.0}{-250.0}{0.0}
\end{figure}
\clearpage

\begin{figure}
\plotfiddle{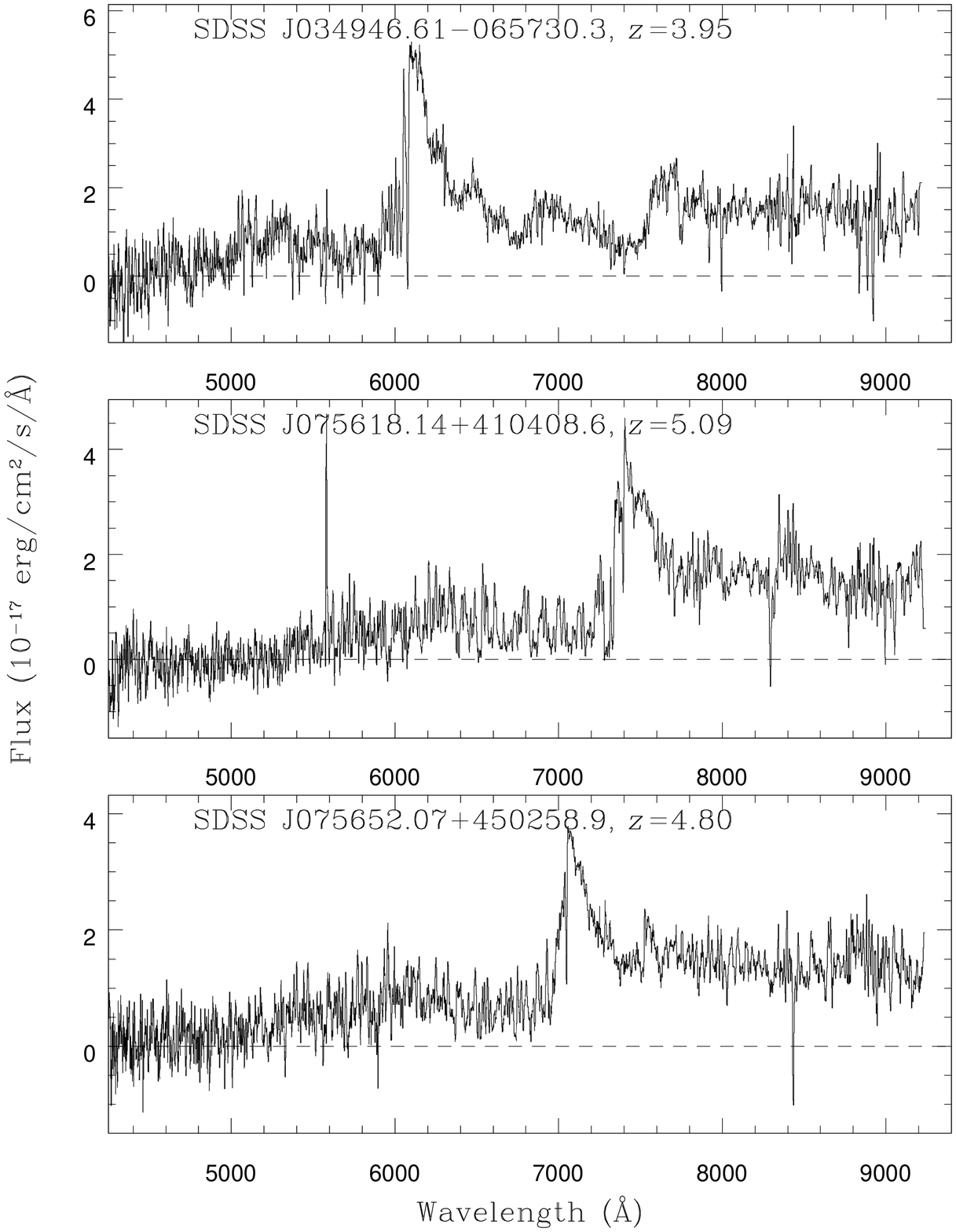}{8.0in}{0.0}{80.0}{80.0}{-250.0}{0.0}
\end{figure}
\clearpage

\begin{figure}
\plotfiddle{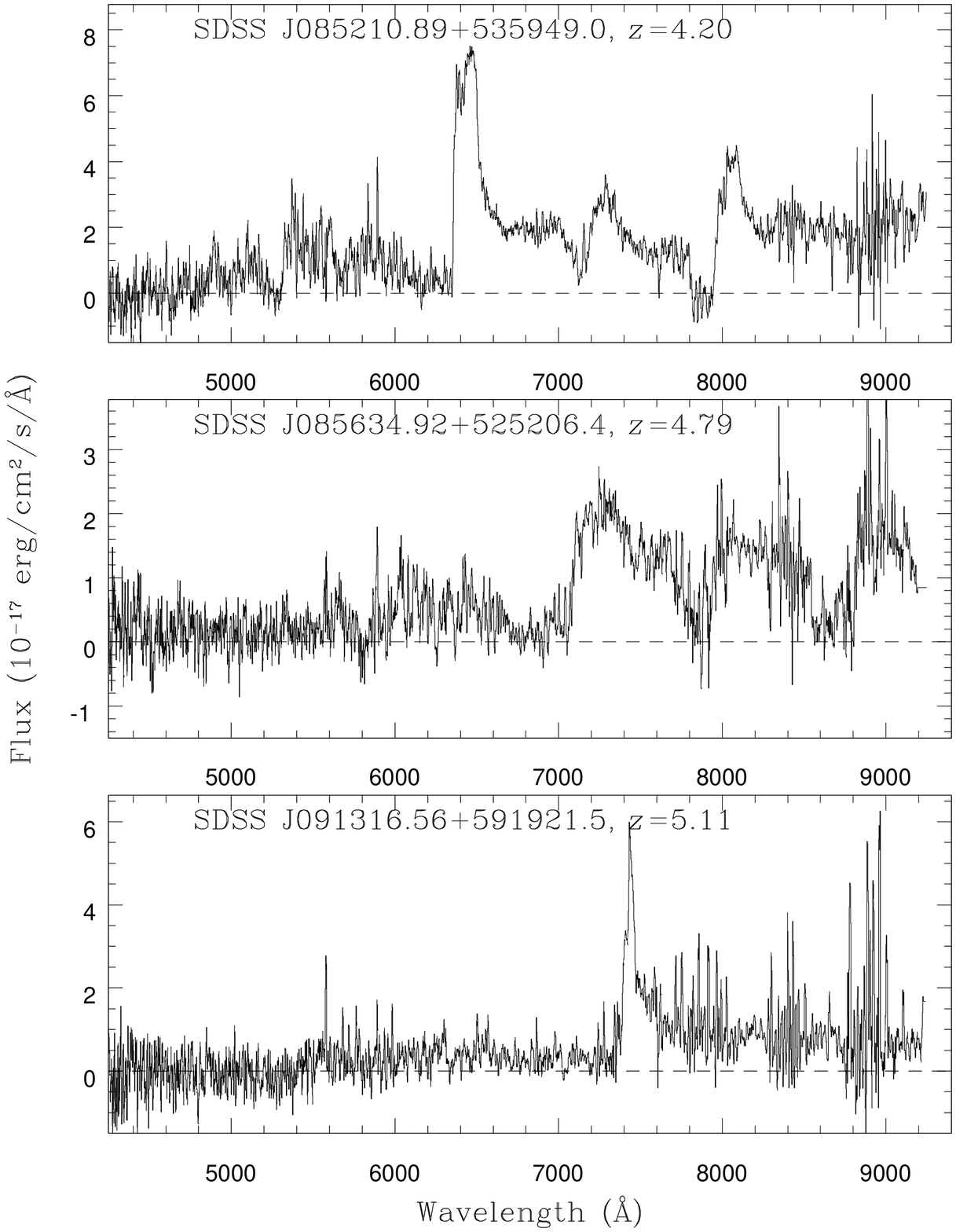}{8.0in}{0.0}{80.0}{80.0}{-250.0}{0.0}
\end{figure}
\clearpage

\begin{figure}
\plotfiddle{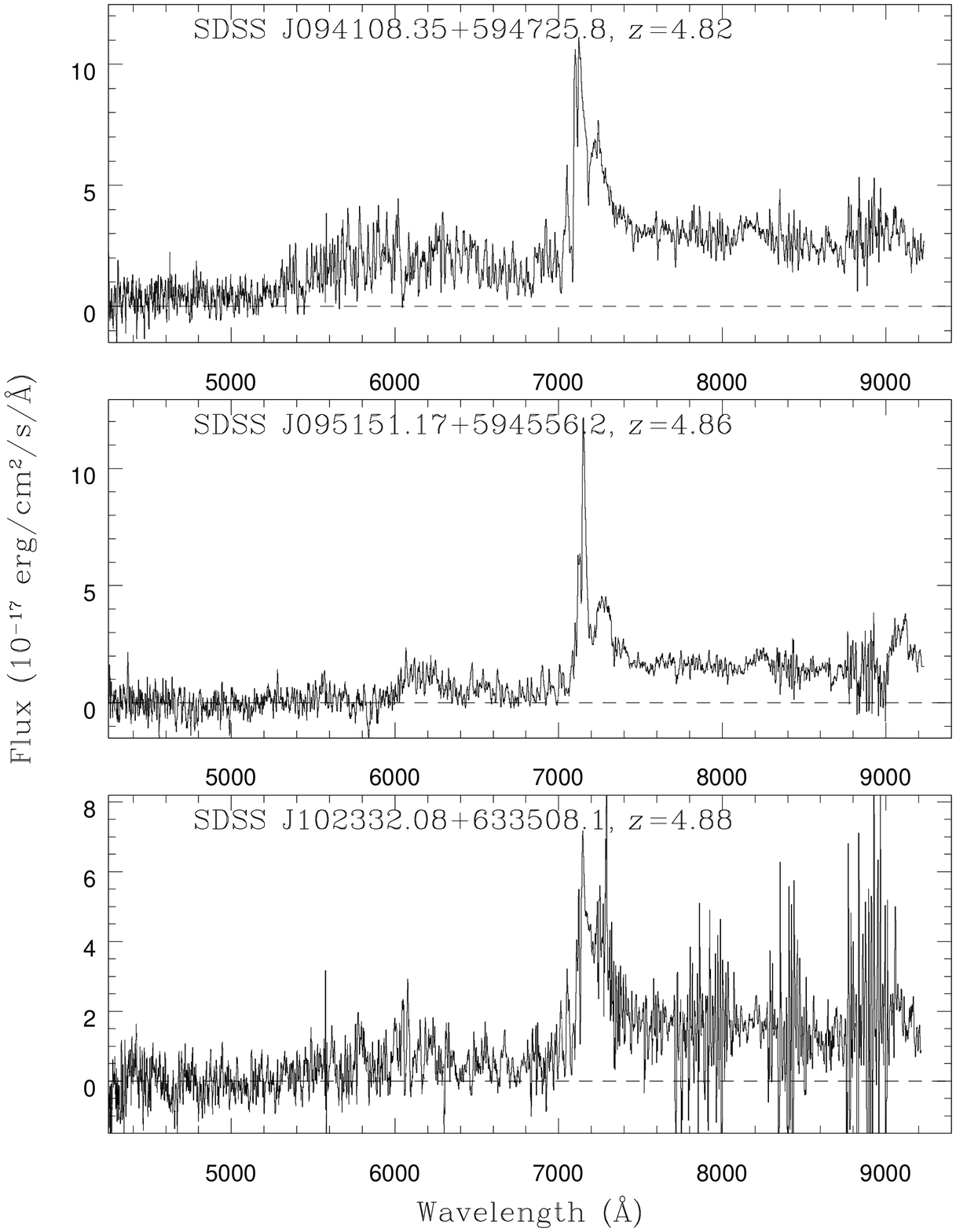}{8.0in}{0.0}{80.0}{80.0}{-250.0}{0.0}
\end{figure}
\clearpage

\begin{figure}
\plotfiddle{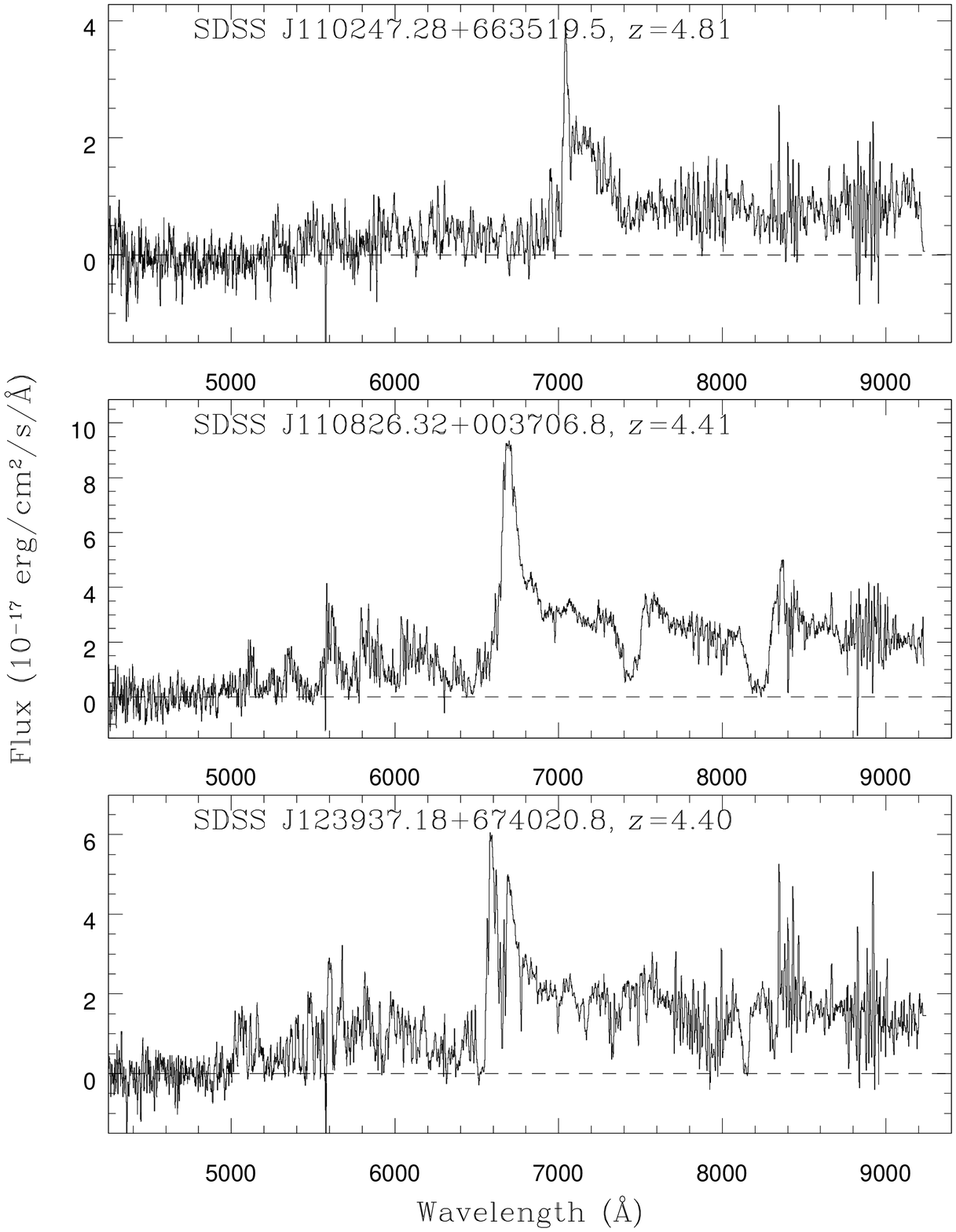}{8.0in}{0.0}{80.0}{80.0}{-250.0}{0.0}
\end{figure}
\clearpage

\begin{figure}
\plotfiddle{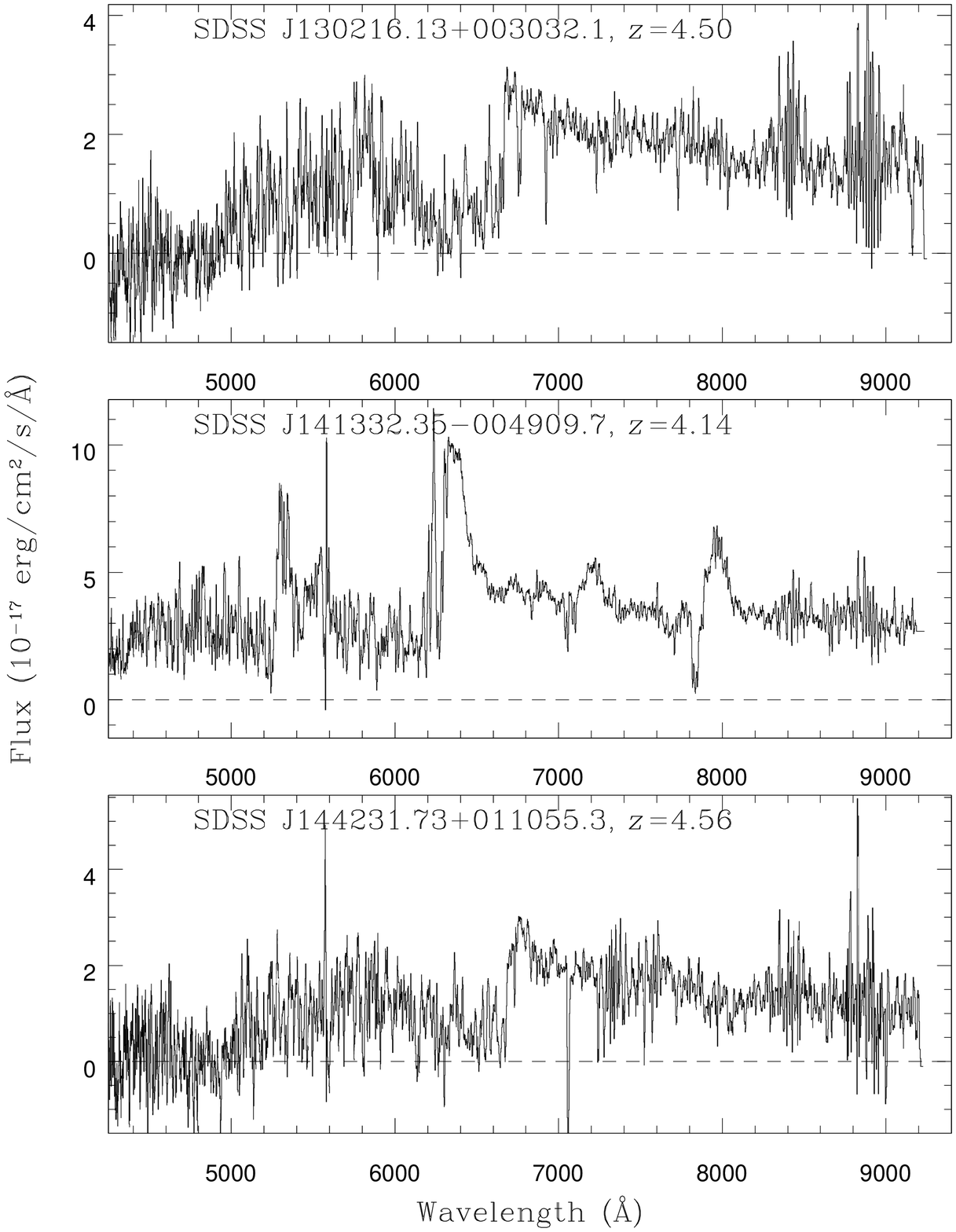}{8.0in}{0.0}{80.0}{80.0}{-250.0}{0.0}
\end{figure}
\clearpage

\begin{figure}
\plotfiddle{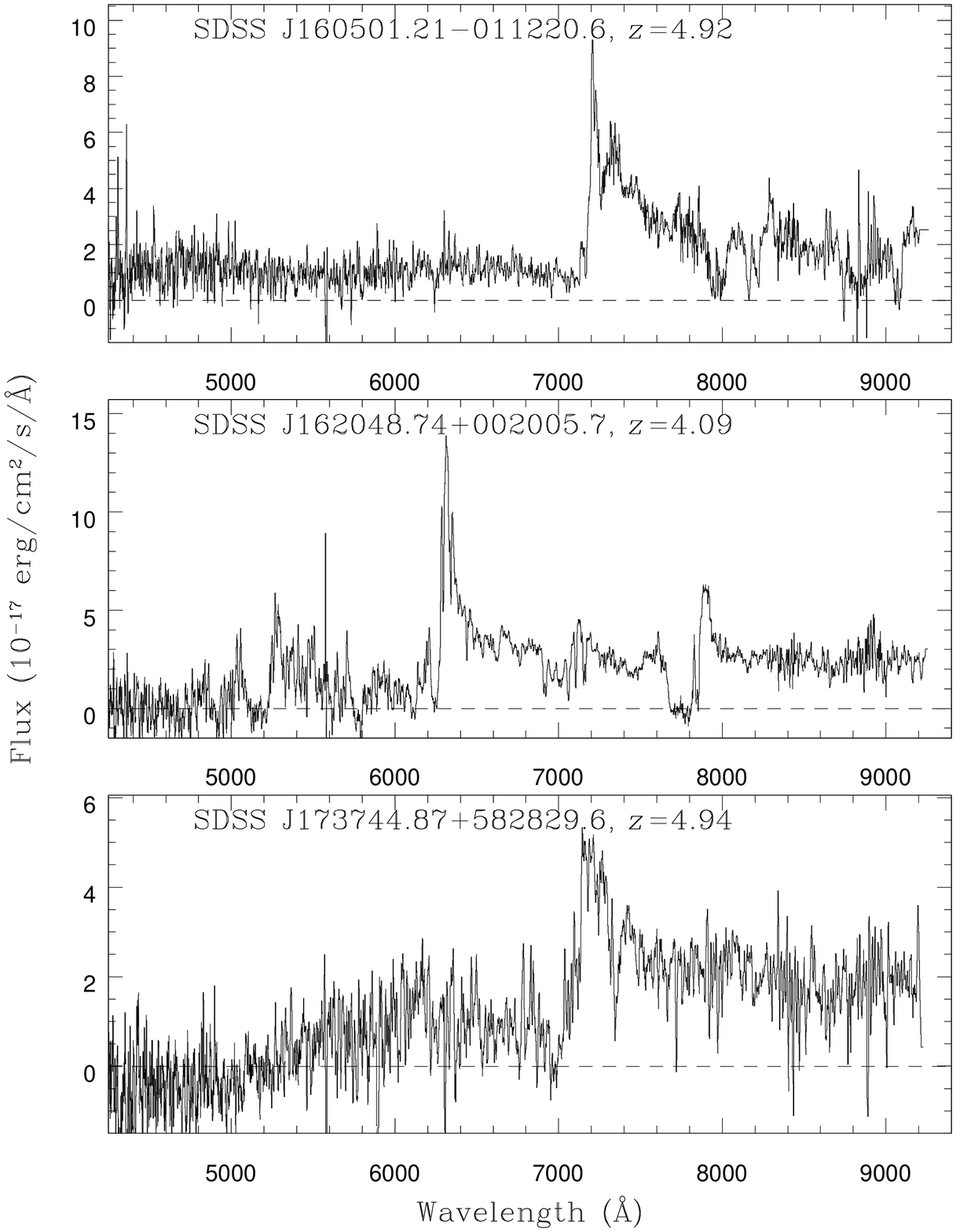}{8.0in}{0.0}{80.0}{80.0}{-250.0}{0.0}
\end{figure}
\clearpage

\begin{figure}
\plotfiddle{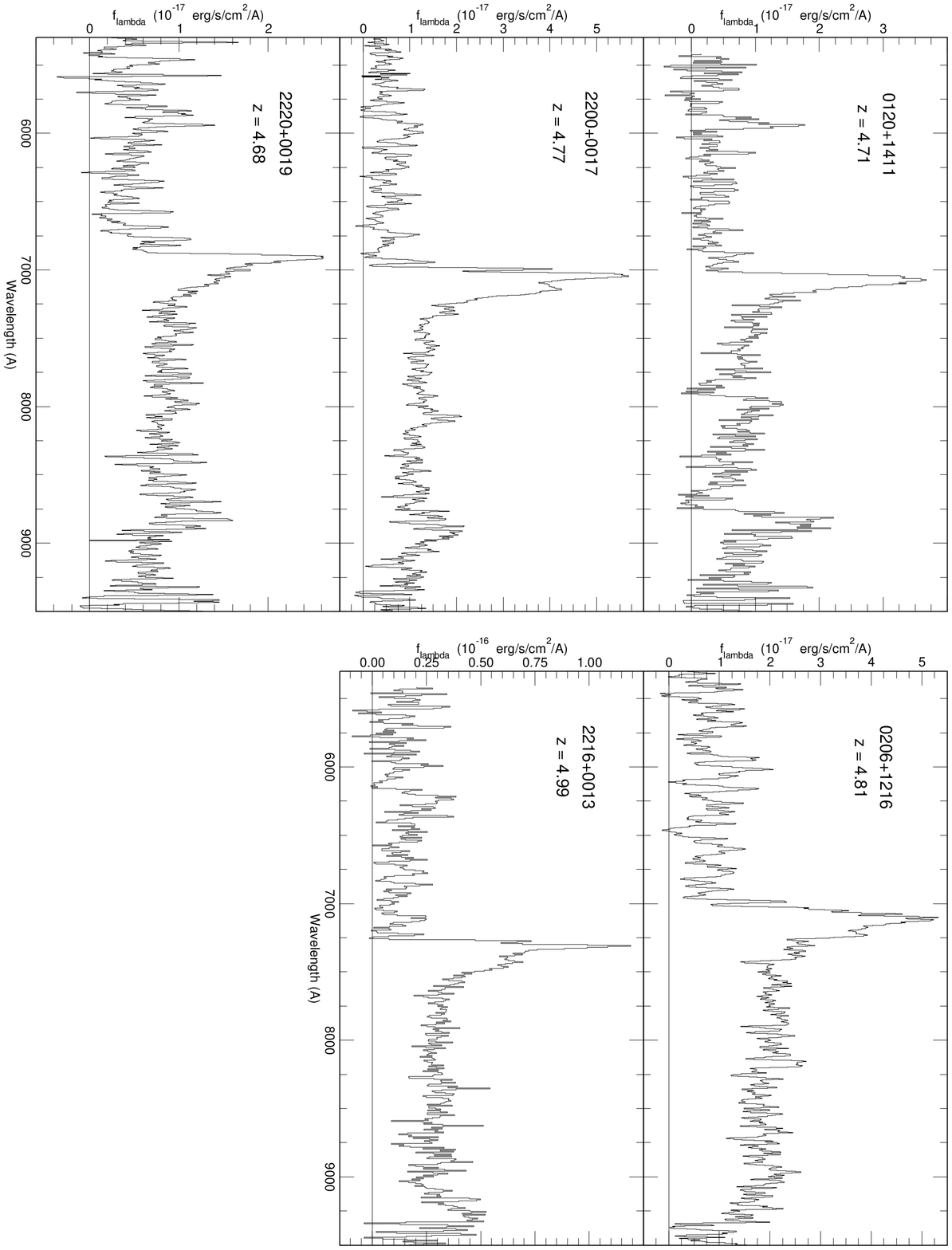}{8.0in}{180.0}{90.0}{90.0}{275.0}{620.0}
\end{figure}
\clearpage

\begin{figure}
\plotfiddle{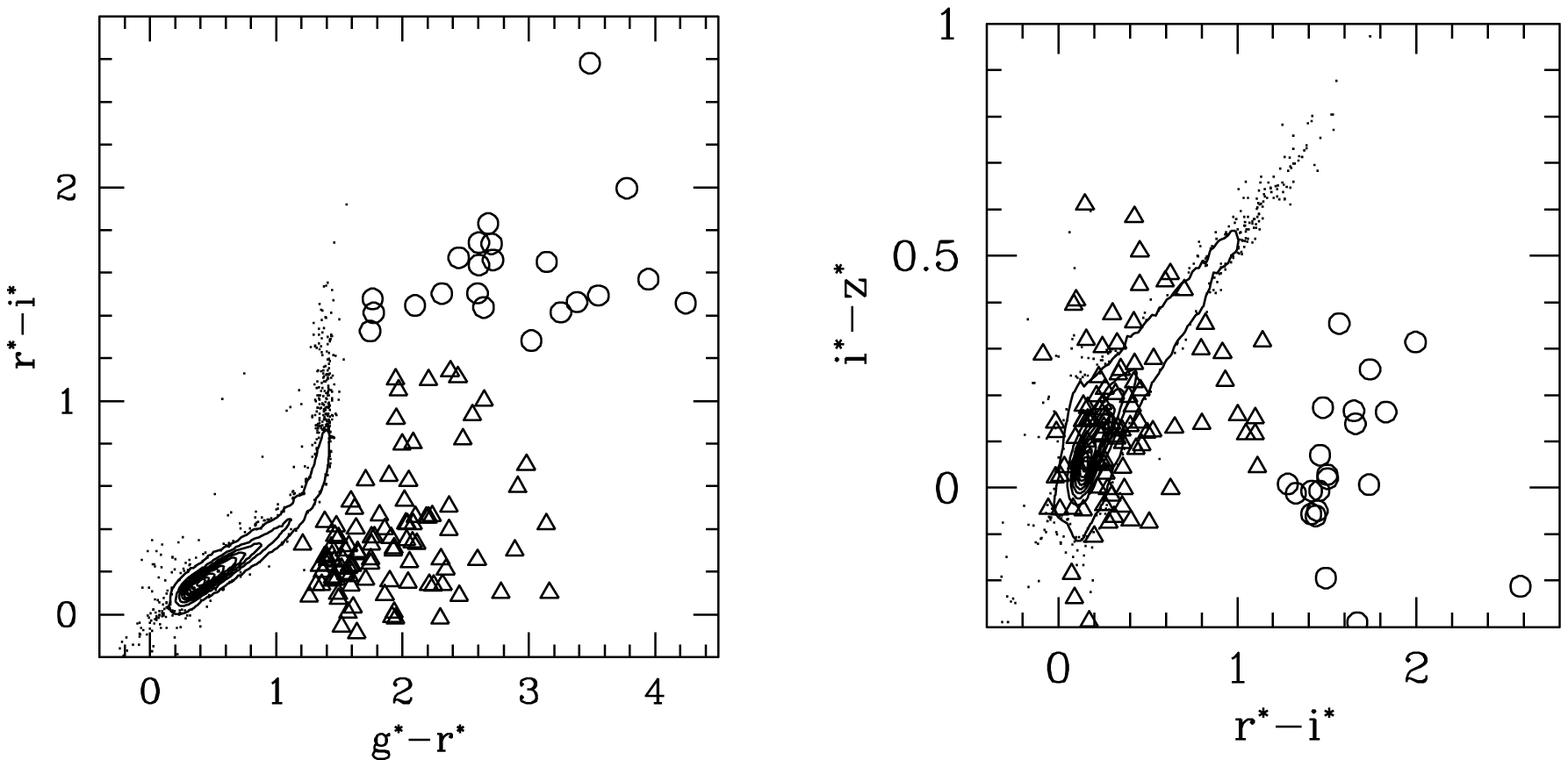}{8.0in}{90.0}{130.0}{130.0}{670.0}{-120.0}
\end{figure}
\clearpage

\begin{figure}
\plotfiddle{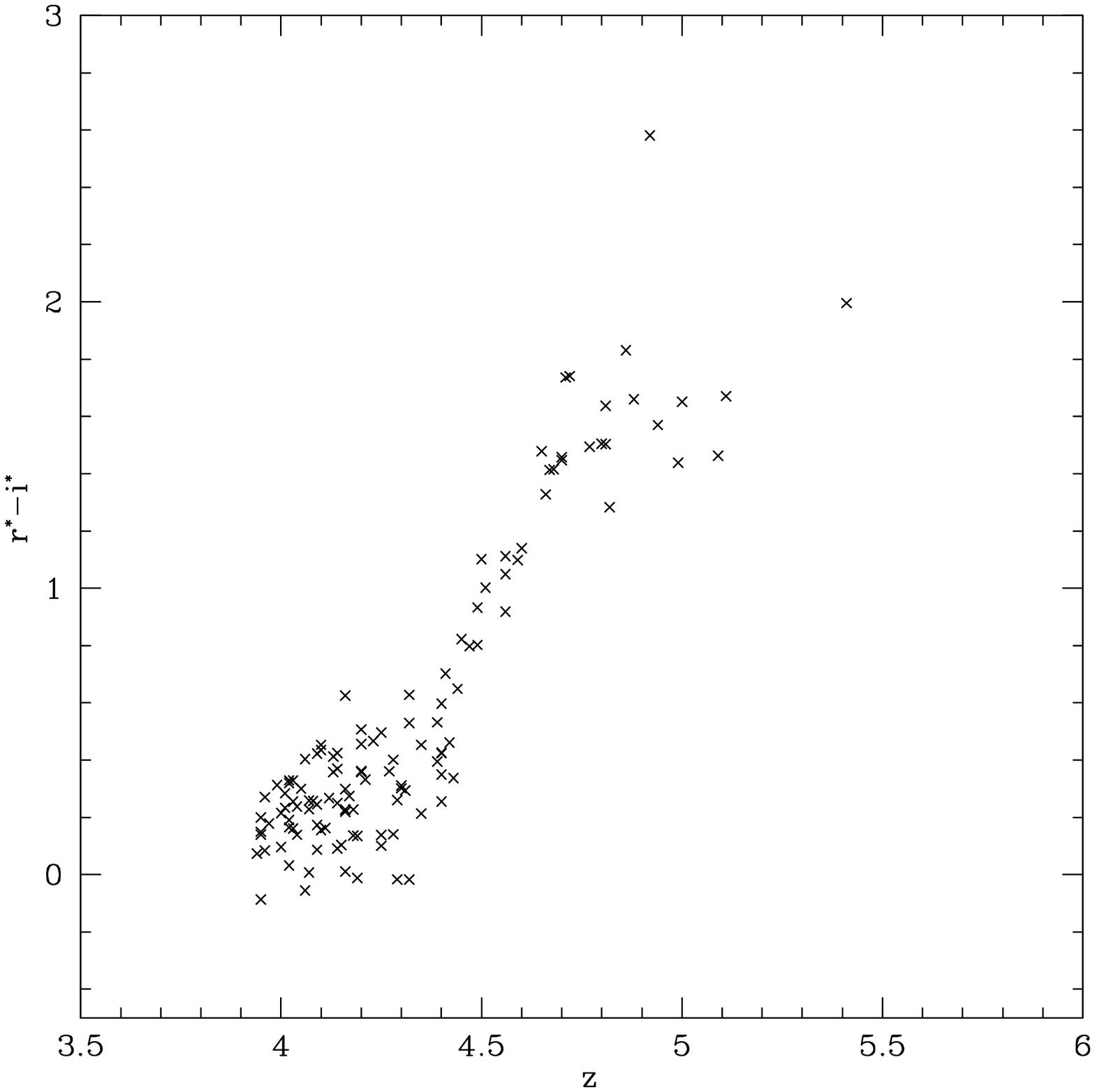}{8.0in}{90.0}{100.0}{100.0}{425.0}{-20.0}
\end{figure}
\clearpage

\begin{figure}
\plotfiddle{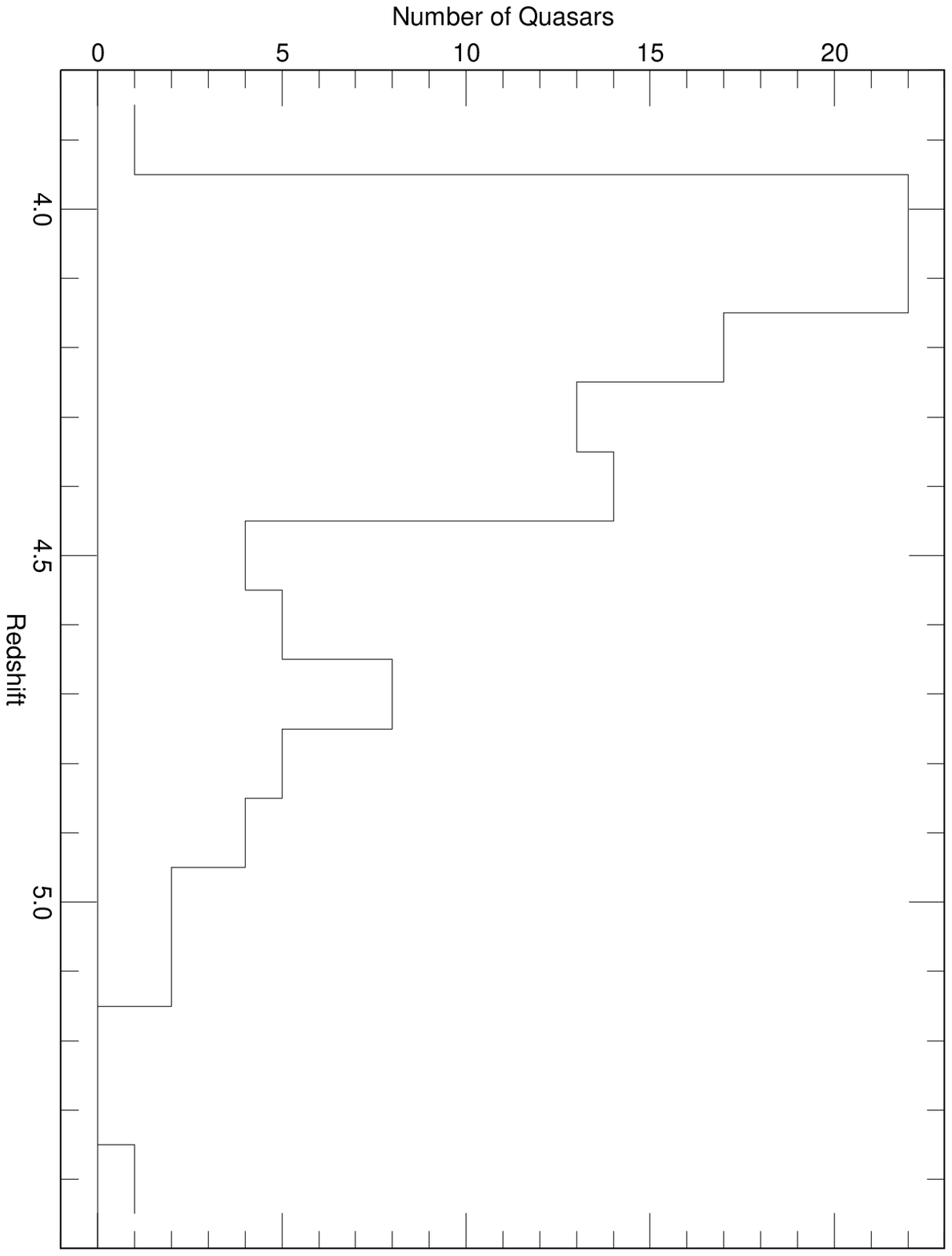}{6.0in}{180.0}{120.0}{120.0}{275.0}{780.0}
\end{figure}
\clearpage

\begin{figure}
\plotfiddle{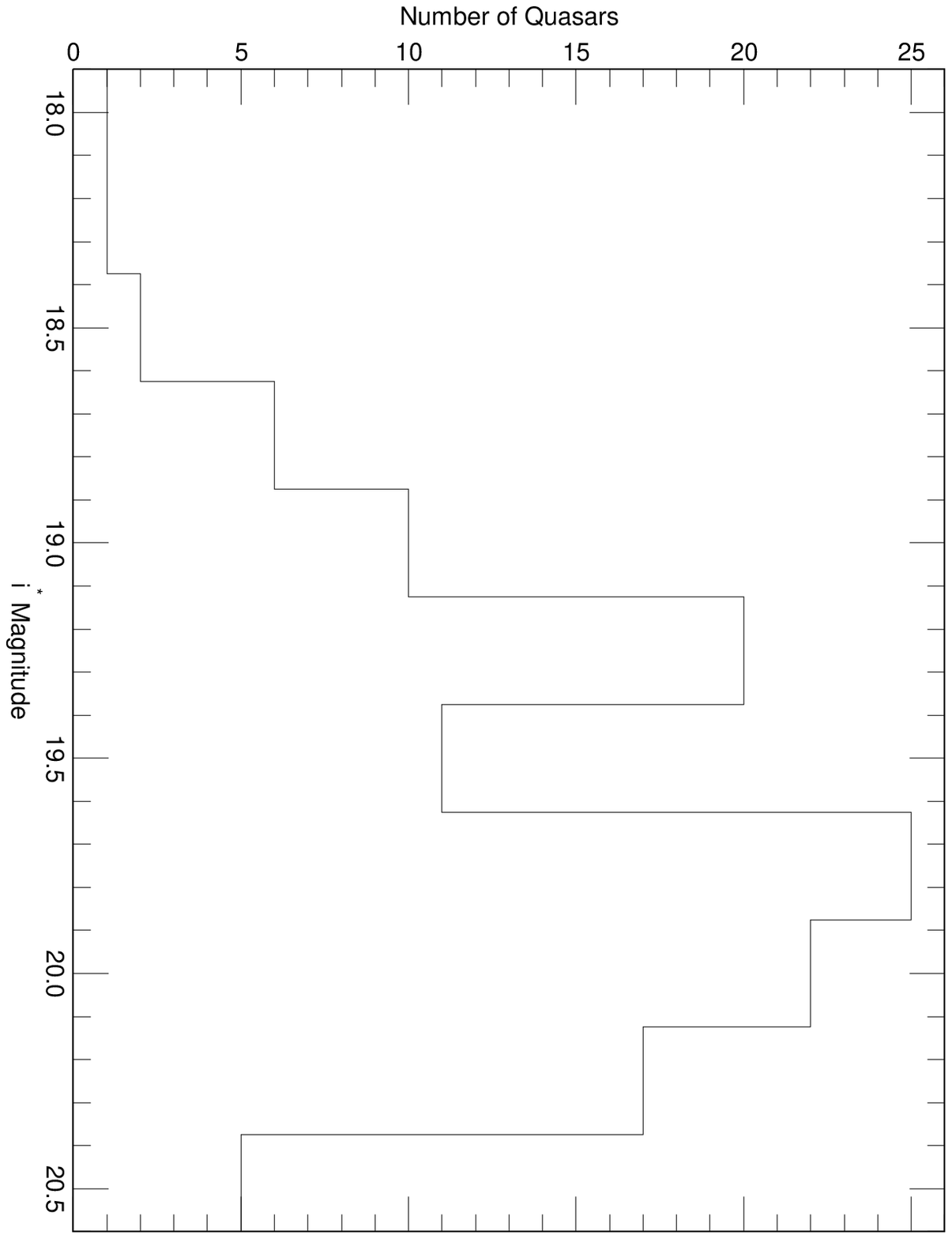}{6.0in}{180.0}{120.0}{120.0}{275.0}{780.0}
\end{figure}
\clearpage

\begin{scriptsize}

\halign{\hskip 12pt
# \hfil \tabskip=1em plus1em minus1em&
\hfil # \hfil &
\hfil # &
\hfil # &
\hfil # &
\hfil # &
\hfil # &
\hfil # \cr
\multispan8{\hfil TABLE 1. Positions and Photometry of SDSS
 High-Redshift Quasars \hfil}\cr
\noalign{\bigskip\hrule\smallskip\hrule\medskip}
\hfil Quasar (SDSSp J) \hfil&\hfil Redshift \hfil 
&\hfil $u^*$ \hfil&\hfil $g^*$ \hfil &
\hfil $r^*$ \hfil & \hfil $i^*$ \hfil & \hfil $z^*$ \hfil & \hfil Note \hfil \cr
\noalign{\medskip\hrule\bigskip}
001950.05$-$004040.8 &  4.31 $\pm$ 0.01
 & 24.17 $\pm$ 0.92 & 21.49 $\pm$ 0.07 & 19.82 $\pm$ 0.02 & 19.50 $\pm$ 0.02 & 19.35 $\pm$ 0.07
 & 8 \cr
003126.80$+$150739.6 &  4.20 $\pm$ 0.01
 & 22.95 $\pm$ 0.44 & 22.32 $\pm$ 0.11 & 20.36 $\pm$ 0.03 & 19.97 $\pm$ 0.03 & 19.81 $\pm$ 0.11
 &   \cr
003749.19$+$155208.4 &  4.05 $\pm$ 0.01
 & 23.79 $\pm$ 0.56 & 21.66 $\pm$ 0.06 & 20.06 $\pm$ 0.02 & 19.72 $\pm$ 0.03 & 19.71 $\pm$ 0.10
 &   \cr
005006.35$-$005319.3 &  4.32 $\pm$ 0.01
 & 23.65 $\pm$ 0.58 & 21.93 $\pm$ 0.09 & 20.18 $\pm$ 0.03 & 19.53 $\pm$ 0.02 & 19.50 $\pm$ 0.09
 &   \cr
005922.65$+$000301.4 &  4.15 $\pm$ 0.01
 & 23.68 $\pm$ 0.78 & 22.45 $\pm$ 0.16 & 19.26 $\pm$ 0.01 & 19.14 $\pm$ 0.02 & 19.07 $\pm$ 0.07
 & 3 \cr
\noalign{\smallskip}
010619.24$+$004823.4 &  4.43 $\pm$ 0.01
 & 23.78 $\pm$ 0.60 & 21.16 $\pm$ 0.04 & 19.05 $\pm$ 0.01 & 18.70 $\pm$ 0.01 & 18.44 $\pm$ 0.03
 & 4 \cr
012004.83$+$141108.3*&  4.71 $\pm$ 0.02
 & 23.48 $\pm$ 0.51 & 24.72 $\pm$ 0.51 & 21.98 $\pm$ 0.10 & 20.22 $\pm$ 0.03 & 20.19 $\pm$ 0.11
 & 1 \cr
012019.99$+$000735.6 &  4.09 $\pm$ 0.02
 & 23.77 $\pm$ 0.82 & 21.51 $\pm$ 0.06 & 20.02 $\pm$ 0.03 & 19.82 $\pm$ 0.03 & 20.08 $\pm$ 0.16
 & 8 \cr
015032.87$+$143425.6*&  4.14 $\pm$ 0.02
 & 24.21 $\pm$ 0.69 & 23.76 $\pm$ 0.32 & 20.56 $\pm$ 0.04 & 20.09 $\pm$ 0.04 & 19.47 $\pm$ 0.08
 &   \cr
015339.61$-$001104.9 &  4.19 $\pm$ 0.01
 & 23.77 $\pm$ 0.55 & 21.17 $\pm$ 0.04 & 18.90 $\pm$ 0.01 & 18.74 $\pm$ 0.01 & 18.58 $\pm$ 0.04
 & 4 \cr
\noalign{\smallskip}
015642.11$+$141944.4*&  4.30 $\pm$ 0.02
 & 23.90 $\pm$ 0.64 & 23.61 $\pm$ 0.28 & 20.67 $\pm$ 0.04 & 20.33 $\pm$ 0.04 & 19.92 $\pm$ 0.09
 &   \cr
015704.10$+$122858.3 &  4.16 $\pm$ 0.02
 & 23.84 $\pm$ 0.66 & 22.24 $\pm$ 0.13 & 20.24 $\pm$ 0.03 & 19.90 $\pm$ 0.03 & 19.72 $\pm$ 0.09
 &   \cr
020651.37$+$121624.4 &  4.81 $\pm$ 0.02
 & 25.13 $\pm$ 0.46 & 24.05 $\pm$ 0.46 & 21.51 $\pm$ 0.07 & 19.86 $\pm$ 0.03 & 19.70 $\pm$ 0.08
 & 1 \cr
021043.15$-$001818.5 &  4.70 $\pm$ 0.02
 & 23.59 $\pm$ 0.67 & 22.87 $\pm$ 0.16 & 20.74 $\pm$ 0.04 & 19.28 $\pm$ 0.02 & 19.31 $\pm$ 0.06
 & 8 \cr
023137.65$-$072854.5 &  5.41 $\pm$ 0.02
 & 24.54 $\pm$ 1.20 & 25.36 $\pm$ 0.63 & 21.54 $\pm$ 0.08 & 19.53 $\pm$ 0.02 & 19.19 $\pm$ 0.07
 &   \cr
\noalign{\smallskip}
023923.47$-$081005.1*&  4.00 $\pm$ 0.02
 & 24.75 $\pm$ 1.31 & 20.90 $\pm$ 0.03 & 19.38 $\pm$ 0.02 & 19.27 $\pm$ 0.02 & 19.14 $\pm$ 0.07
 &   \cr
024447.79$-$081606.1 &  4.03 $\pm$ 0.02
 & 24.20 $\pm$ 1.48 & 19.69 $\pm$ 0.02 & 18.21 $\pm$ 0.02 & 18.03 $\pm$ 0.01 & 17.92 $\pm$ 0.03
 &   \cr
024457.19$-$010809.9*&  3.96 $\pm$ 0.02
 & 22.81 $\pm$ 0.38 & 20.04 $\pm$ 0.02 & 18.62 $\pm$ 0.01 & 18.33 $\pm$ 0.01 & 18.16 $\pm$ 0.02
 & 2 \cr
025039.17$-$065405.1 &  4.45 $\pm$ 0.01
 & 25.00 $\pm$ 1.23 & 23.20 $\pm$ 0.23 & 20.68 $\pm$ 0.04 & 19.84 $\pm$ 0.03 & 19.46 $\pm$ 0.09
 &   \cr
025159.41$-$084258.1 &  4.25 $\pm$ 0.01
 & 23.06 $\pm$ 0.82 & 23.52 $\pm$ 0.32 & 20.69 $\pm$ 0.04 & 20.56 $\pm$ 0.05 & 20.13 $\pm$ 0.14
 &   \cr
\noalign{\smallskip}
025204.29$+$003136.9*&  4.10 $\pm$ 0.02
 & 24.88 $\pm$ 0.66 & 22.43 $\pm$ 0.14 & 20.15 $\pm$ 0.03 & 19.64 $\pm$ 0.03 & 19.15 $\pm$ 0.06
 &   \cr
025518.58$+$004847.6 &  3.99 $\pm$ 0.01
 & 23.92 $\pm$ 0.61 & 20.59 $\pm$ 0.02 & 19.07 $\pm$ 0.01 & 18.70 $\pm$ 0.01 & 18.71 $\pm$ 0.04
 &   \cr
025647.06$-$085041.4*&  4.21 $\pm$ 0.02
 & 24.08 $\pm$ 0.73 & 22.14 $\pm$ 0.12 & 19.98 $\pm$ 0.02 & 19.62 $\pm$ 0.03 & 19.46 $\pm$ 0.08
 &   \cr
030025.23$+$003224.3 &  4.19 $\pm$ 0.01
 & 22.76 $\pm$ 0.32 & 22.01 $\pm$ 0.09 & 19.99 $\pm$ 0.02 & 19.94 $\pm$ 0.03 & 19.76 $\pm$ 0.09
 & 8 \cr
031213.98$-$062658.8 &  4.03 $\pm$ 0.01
 & 24.89 $\pm$ 1.26 & 20.71 $\pm$ 0.03 & 19.44 $\pm$ 0.02 & 19.08 $\pm$ 0.02 & 18.93 $\pm$ 0.05
 &   \cr
\noalign{\smallskip}
032226.10$-$055824.7 &  3.94 $\pm$ 0.01
 & 25.87 $\pm$ 0.85 & 21.56 $\pm$ 0.06 & 20.01 $\pm$ 0.03 & 19.90 $\pm$ 0.04 & 20.06 $\pm$ 0.14
 &   \cr
032608.12$-$003340.1 &  4.18 $\pm$ 0.01
 & 23.48 $\pm$ 0.61 & 21.17 $\pm$ 0.04 & 19.47 $\pm$ 0.01 & 19.26 $\pm$ 0.02 & 19.13 $\pm$ 0.06
 & 6 \cr
033119.67$-$074143.1 &  4.70 $\pm$ 0.01
 & 24.90 $\pm$ 1.19 & 24.93 $\pm$ 0.84 & 20.61 $\pm$ 0.04 & 19.10 $\pm$ 0.02 & 19.07 $\pm$ 0.07
 &   \cr
033305.32$-$053709.0*&  4.09 $\pm$ 0.01
 & 24.26 $\pm$ 1.19 & 22.40 $\pm$ 0.10 & 19.90 $\pm$ 0.03 & 19.79 $\pm$ 0.04 & 19.37 $\pm$ 0.07
 &   \cr
033344.43$-$060625.2 &  4.09 $\pm$ 0.01
 & 23.35 $\pm$ 0.64 & 22.80 $\pm$ 0.15 & 20.70 $\pm$ 0.04 & 20.43 $\pm$ 0.05 & 20.10 $\pm$ 0.13
 &   \cr
\noalign{\smallskip}
033406.99$-$063406.5 &  3.95 $\pm$ 0.01
 & 25.55 $\pm$ 0.89 & 20.75 $\pm$ 0.03 & 19.35 $\pm$ 0.02 & 19.18 $\pm$ 0.02 & 19.21 $\pm$ 0.07
 &   \cr
033829.31$+$002156.3 &  5.00 $\pm$ 0.02
 & 23.70 $\pm$ 0.61 & 24.92 $\pm$ 0.51 & 21.68 $\pm$ 0.09 & 19.97 $\pm$ 0.03 & 19.75 $\pm$ 0.13
 & 6 \cr
034109.35$-$064805.1 &  4.08 $\pm$ 0.01
 & 25.32 $\pm$ 0.88 & 22.07 $\pm$ 0.08 & 20.49 $\pm$ 0.03 & 20.20 $\pm$ 0.04 & 20.20 $\pm$ 0.15
 &   \cr
034541.51$-$072315.3 &  4.02 $\pm$ 0.02
 & 25.30 $\pm$ 0.91 & 21.73 $\pm$ 0.07 & 19.90 $\pm$ 0.02 & 19.52 $\pm$ 0.02 & 19.16 $\pm$ 0.06
 &   \cr
034946.61$-$065730.3*&  3.95 $\pm$ 0.02
 & 23.78 $\pm$ 0.93 & 22.59 $\pm$ 0.13 & 20.44 $\pm$ 0.03 & 20.23 $\pm$ 0.03 & 19.56 $\pm$ 0.08
 &   \cr
\noalign{\smallskip}
073147.01$+$364346.5 &  4.03 $\pm$ 0.01
 & 24.48 $\pm$ 0.98 & 21.05 $\pm$ 0.04 & 19.58 $\pm$ 0.03 & 19.29 $\pm$ 0.02 & 19.20 $\pm$ 0.06
 &   \cr
075103.96$+$424211.6 &  4.16 $\pm$ 0.01
 & 23.79 $\pm$ 0.72 & 20.69 $\pm$ 0.03 & 19.07 $\pm$ 0.02 & 18.82 $\pm$ 0.02 & 18.80 $\pm$ 0.04
 &   \cr
075618.14$+$410408.6 &  5.09 $\pm$ 0.01
 & 24.53 $\pm$ 0.91 & 25.13 $\pm$ 0.62 & 21.70 $\pm$ 0.08 & 20.21 $\pm$ 0.03 & 20.11 $\pm$ 0.12
 &   \cr
075652.07$+$450258.9 &  4.80 $\pm$ 0.01
 & 24.47 $\pm$ 0.95 & 24.41 $\pm$ 0.49 & 21.78 $\pm$ 0.09 & 20.24 $\pm$ 0.04 & 20.20 $\pm$ 0.13
 &   \cr
075732.89$+$441424.7 &  4.16 $\pm$ 0.02
 & 23.70 $\pm$ 0.48 & 21.02 $\pm$ 0.03 & 19.63 $\pm$ 0.02 & 19.38 $\pm$ 0.02 & 19.30 $\pm$ 0.06
 &   \cr
\noalign{\smallskip}
080159.25$+$433625.0 &  4.17 $\pm$ 0.01
 & 24.07 $\pm$ 0.80 & 22.04 $\pm$ 0.08 & 20.56 $\pm$ 0.03 & 20.26 $\pm$ 0.03 & 20.24 $\pm$ 0.10
 &   \cr
080549.94$+$482345.9 &  4.18 $\pm$ 0.03
 & 25.63 $\pm$ 0.74 & 21.76 $\pm$ 0.07 & 20.11 $\pm$ 0.03 & 19.86 $\pm$ 0.03 & 19.59 $\pm$ 0.07
 &   \cr
081054.88$+$460357.9 &  4.07 $\pm$ 0.01
 & 25.45 $\pm$ 0.88 & 20.46 $\pm$ 0.03 & 18.67 $\pm$ 0.01 & 18.38 $\pm$ 0.02 & 18.30 $\pm$ 0.03
 &   \cr
081241.12$+$442129.0 &  4.32 $\pm$ 0.01
 & 22.91 $\pm$ 0.50 & 21.98 $\pm$ 0.08 & 19.97 $\pm$ 0.03 & 19.95 $\pm$ 0.04 & 19.77 $\pm$ 0.08
 &   \cr
083046.28$+$474646.5 &  4.29 $\pm$ 0.01
 & 25.40 $\pm$ 0.73 & 23.03 $\pm$ 0.22 & 20.69 $\pm$ 0.04 & 20.40 $\pm$ 0.04 & 20.16 $\pm$ 0.11
 &   \cr
\noalign{\medskip\hrule}}

\clearpage

\halign{\hskip 12pt
# \hfil \tabskip=1em plus1em minus1em&
\hfil # \hfil &
\hfil # &
\hfil # &
\hfil # &
\hfil # &
\hfil # &
\hfil # \cr
\multispan8{\hfil TABLE 1. Positions and Photometry of SDSS
 High-Redshift Quasars (continued) \hfil}\cr
\noalign{\bigskip\hrule\smallskip\hrule\medskip}
\hfil Quasar (SDSSp J) \hfil&\hfil Redshift \hfil 
&\hfil $u^*$ \hfil&\hfil $g^*$ \hfil &
\hfil $r^*$ \hfil & \hfil $i^*$ \hfil & \hfil $z^*$ \hfil & \hfil Note \hfil \cr
\noalign{\medskip\hrule\bigskip}
083103.00$+$523533.6 &  4.44 $\pm$ 0.01
 & 23.83 $\pm$ 0.89 & 22.00 $\pm$ 0.08 & 20.05 $\pm$ 0.02 & 19.37 $\pm$ 0.02 & 19.22 $\pm$ 0.07
 &   \cr
083212.37$+$530327.4 &  4.02 $\pm$ 0.01
 & 23.60 $\pm$ 0.74 & 21.44 $\pm$ 0.05 & 19.82 $\pm$ 0.03 & 19.46 $\pm$ 0.03 & 19.23 $\pm$ 0.06
 &   \cr
083324.57$+$523955.0 &  3.97 $\pm$ 0.01
 & 25.03 $\pm$ 0.91 & 21.76 $\pm$ 0.06 & 20.14 $\pm$ 0.03 & 19.94 $\pm$ 0.03 & 19.77 $\pm$ 0.09
 &   \cr
083946.22$+$511202.8 &  4.39 $\pm$ 0.01
 & 25.22 $\pm$ 0.78 & 21.78 $\pm$ 0.06 & 19.39 $\pm$ 0.02 & 18.98 $\pm$ 0.02 & 18.77 $\pm$ 0.04
 &   \cr
084811.52$-$001418.0 &  4.12 $\pm$ 0.01
 & 23.27 $\pm$ 0.42 & 20.71 $\pm$ 0.03 & 19.28 $\pm$ 0.01 & 18.99 $\pm$ 0.01 & 18.94 $\pm$ 0.04
 &   \cr
\noalign{\smallskip}
085151.27$+$020756.1 &  4.28 $\pm$ 0.01
 & 24.22 $\pm$ 0.96 & 21.70 $\pm$ 0.06 & 19.45 $\pm$ 0.02 & 19.29 $\pm$ 0.02 & 19.09 $\pm$ 0.06
 &   \cr
085210.89$+$535949.0*&  4.20 $\pm$ 0.02
 & 23.75 $\pm$ 0.63 & 22.42 $\pm$ 0.10 & 20.29 $\pm$ 0.03 & 19.82 $\pm$ 0.03 & 19.29 $\pm$ 0.05
 &   \cr
085430.18$+$004213.6 &  4.07 $\pm$ 0.01
 & 24.25 $\pm$ 0.73 & 21.91 $\pm$ 0.07 & 20.28 $\pm$ 0.03 & 20.02 $\pm$ 0.04 & 19.82 $\pm$ 0.09
 &   \cr
085634.93$+$525206.4*&  4.79 $\pm$ 0.10
 & 25.16 $\pm$ 0.92 & 24.69 $\pm$ 0.50 & 22.06 $\pm$ 0.11 & 20.31 $\pm$ 0.04 & 20.04 $\pm$ 0.09
 &   \cr
090242.08$-$002125.9 &  4.39 $\pm$ 0.02
 & 25.43 $\pm$ 0.68 & 22.38 $\pm$ 0.09 & 20.33 $\pm$ 0.03 & 19.77 $\pm$ 0.02 & 19.47 $\pm$ 0.06
 &   \cr
\noalign{\smallskip}
090440.64$+$535038.8 &  4.25 $\pm$ 0.02
 & 23.83 $\pm$ 0.77 & 21.40 $\pm$ 0.04 & 19.76 $\pm$ 0.02 & 19.25 $\pm$ 0.04 & 19.13 $\pm$ 0.04
 &   \cr
090532.15$-$001430.5 &  4.25 $\pm$ 0.01
 & 25.01 $\pm$ 0.77 & 22.42 $\pm$ 0.10 & 20.05 $\pm$ 0.03 & 19.89 $\pm$ 0.02 & 19.77 $\pm$ 0.08
 &   \cr
091016.79$+$575331.1 &  4.01 $\pm$ 0.03
 & 24.73 $\pm$ 1.31 & 21.51 $\pm$ 0.05 & 19.81 $\pm$ 0.03 & 19.50 $\pm$ 0.03 & 19.55 $\pm$ 0.07
 &   \cr
091316.56$+$591921.5 &  5.11 $\pm$ 0.02
 & 24.24 $\pm$ 1.06 & 24.68 $\pm$ 0.55 & 22.19 $\pm$ 0.14 & 20.50 $\pm$ 0.05 & 20.76 $\pm$ 0.25
 &   \cr
092038.49$+$564235.9 &  4.14 $\pm$ 0.01
 & 24.47 $\pm$ 1.01 & 21.85 $\pm$ 0.06 & 20.34 $\pm$ 0.03 & 19.95 $\pm$ 0.03 & 19.92 $\pm$ 0.12
 &   \cr
\noalign{\smallskip}
092256.20$+$561849.3 &  4.19 $\pm$ 0.02
 & 23.83 $\pm$ 0.90 & 21.17 $\pm$ 0.04 & 19.37 $\pm$ 0.03 & 18.99 $\pm$ 0.02 & 18.92 $\pm$ 0.04
 &   \cr
094056.02$+$584830.2 &  4.66 $\pm$ 0.01
 & 24.19 $\pm$ 0.96 & 22.37 $\pm$ 0.10 & 20.61 $\pm$ 0.04 & 19.27 $\pm$ 0.02 & 19.27 $\pm$ 0.07
 &   \cr
094108.36$+$594725.8 &  4.82 $\pm$ 0.02
 & 23.91 $\pm$ 0.77 & 23.70 $\pm$ 0.33 & 20.66 $\pm$ 0.04 & 19.36 $\pm$ 0.03 & 19.34 $\pm$ 0.06
 &   \cr
094917.17$+$602104.5 &  4.28 $\pm$ 0.01
 & 24.07 $\pm$ 0.89 & 22.55 $\pm$ 0.14 & 20.67 $\pm$ 0.04 & 20.26 $\pm$ 0.04 & 20.12 $\pm$ 0.11
 &   \cr
095000.17$+$620318.6 &  4.06 $\pm$ 0.01
 & 25.81 $\pm$ 0.67 & 21.73 $\pm$ 0.07 & 20.19 $\pm$ 0.03 & 20.22 $\pm$ 0.04 & 20.25 $\pm$ 0.17
 &   \cr
\noalign{\smallskip}
095151.17$+$594556.2 &  4.86 $\pm$ 0.03
 & 23.53 $\pm$ 0.71 & 24.34 $\pm$ 0.48 & 21.65 $\pm$ 0.08 & 19.81 $\pm$ 0.03 & 19.64 $\pm$ 0.08
 &   \cr
100154.87$+$630818.0 &  4.14 $\pm$ 0.01
 & 25.09 $\pm$ 0.96 & 22.31 $\pm$ 0.09 & 20.43 $\pm$ 0.03 & 20.32 $\pm$ 0.04 & 20.54 $\pm$ 0.18
 &   \cr
100413.14$+$630437.4 &  4.11 $\pm$ 0.02
 & 25.54 $\pm$ 0.77 & 22.15 $\pm$ 0.10 & 20.42 $\pm$ 0.03 & 20.24 $\pm$ 0.04 & 20.08 $\pm$ 0.14
 &   \cr
101053.52$+$644832.0 &  4.65 $\pm$ 0.04
 & 23.90 $\pm$ 0.80 & 23.17 $\pm$ 0.21 & 21.37 $\pm$ 0.06 & 19.87 $\pm$ 0.03 & 19.67 $\pm$ 0.08
 &   \cr
101549.00$+$002020.0 &  4.40 $\pm$ 0.02
 & 23.83 $\pm$ 0.50 & 21.78 $\pm$ 0.06 & 19.72 $\pm$ 0.01 & 19.27 $\pm$ 0.01 & 18.98 $\pm$ 0.04
 & 5 \cr
\noalign{\smallskip}
102043.82$+$000105.8 &  4.16 $\pm$ 0.04
 & 23.15 $\pm$ 0.54 & 22.57 $\pm$ 0.15 & 20.48 $\pm$ 0.03 & 19.83 $\pm$ 0.02 & 19.35 $\pm$ 0.08
 &   \cr
102332.08$+$633508.1 &  4.88 $\pm$ 0.01
 & 23.07 $\pm$ 0.48 & 24.09 $\pm$ 0.47 & 21.36 $\pm$ 0.06 & 19.69 $\pm$ 0.03 & 19.55 $\pm$ 0.07
 &   \cr
103309.21$+$644351.1 &  3.95 $\pm$ 0.02
 & 24.16 $\pm$ 1.04 & 21.47 $\pm$ 0.06 & 20.09 $\pm$ 0.03 & 19.89 $\pm$ 0.03 & 19.98 $\pm$ 0.13
 &   \cr
104008.10$+$651429.3 &  4.49 $\pm$ 0.02
 & 24.05 $\pm$ 0.97 & 23.34 $\pm$ 0.24 & 20.77 $\pm$ 0.04 & 19.83 $\pm$ 0.03 & 19.59 $\pm$ 0.08
 &   \cr
104040.14$-$001540.9 &  4.32 $\pm$ 0.03
 & 23.49 $\pm$ 0.61 & 21.00 $\pm$ 0.04 & 19.36 $\pm$ 0.01 & 18.80 $\pm$ 0.01 & 18.65 $\pm$ 0.03
 &   \cr
\noalign{\smallskip}
104351.20$+$650647.7 &  4.49 $\pm$ 0.02
 & 26.30 $\pm$ 0.55 & 22.00 $\pm$ 0.08 & 19.90 $\pm$ 0.03 & 19.09 $\pm$ 0.02 & 18.95 $\pm$ 0.05
 &   \cr
104837.40$-$002813.7 &  4.00 $\pm$ 0.02
 & 24.71 $\pm$ 0.58 & 20.86 $\pm$ 0.03 & 19.29 $\pm$ 0.02 & 19.05 $\pm$ 0.07 & 18.96 $\pm$ 0.12
 & 9 \cr
105254.60$-$000625.9 &  4.13 $\pm$ 0.01
 & 23.95 $\pm$ 0.46 & 21.47 $\pm$ 0.04 & 19.93 $\pm$ 0.02 & 19.54 $\pm$ 0.02 & 19.55 $\pm$ 0.06
 &   \cr
105320.43$-$001649.6 &  4.29 $\pm$ 0.01
 & 23.69 $\pm$ 0.58 & 21.70 $\pm$ 0.07 & 19.35 $\pm$ 0.01 & 19.33 $\pm$ 0.02 & 19.27 $\pm$ 0.06
 & 5 \cr
105602.37$+$003222.0 &  4.02 $\pm$ 0.01
 & 22.99 $\pm$ 0.47 & 21.47 $\pm$ 0.05 & 19.88 $\pm$ 0.02 & 19.66 $\pm$ 0.02 & 19.49 $\pm$ 0.06
 &   \cr
\noalign{\smallskip}
105902.73$+$010404.1 &  4.06 $\pm$ 0.01
 & 23.69 $\pm$ 0.36 & 21.30 $\pm$ 0.04 & 19.64 $\pm$ 0.01 & 19.21 $\pm$ 0.01 & 19.26 $\pm$ 0.05
 &   \cr
110247.29$+$663519.5 &  4.81 $\pm$ 0.03
 & 24.55 $\pm$ 1.18 & 24.74 $\pm$ 0.53 & 22.12 $\pm$ 0.12 & 20.47 $\pm$ 0.04 & 20.98 $\pm$ 0.32
 &   \cr
110813.86$-$005944.5 &  4.01 $\pm$ 0.01
 & 24.20 $\pm$ 0.48 & 20.96 $\pm$ 0.03 & 19.51 $\pm$ 0.01 & 19.25 $\pm$ 0.02 & 19.11 $\pm$ 0.05
 & 9 \cr
110819.16$-$005824.0 &  4.56 $\pm$ 0.02
 & 23.97 $\pm$ 0.48 & 22.96 $\pm$ 0.18 & 20.94 $\pm$ 0.04 & 19.87 $\pm$ 0.02 & 19.72 $\pm$ 0.08
 &   \cr
110826.32$+$003706.8*&  4.41 $\pm$ 0.02
 & 24.36 $\pm$ 0.73 & 23.58 $\pm$ 0.33 & 20.57 $\pm$ 0.03 & 19.84 $\pm$ 0.03 & 19.40 $\pm$ 0.06
 &   \cr
\noalign{\smallskip}
111224.18$+$004630.4 &  4.02 $\pm$ 0.01
 & 24.32 $\pm$ 0.46 & 21.36 $\pm$ 0.04 & 19.71 $\pm$ 0.01 & 19.66 $\pm$ 0.02 & 19.59 $\pm$ 0.07
 &   \cr
111401.47$-$005321.2 &  4.59 $\pm$ 0.01
 & 23.43 $\pm$ 0.44 & 22.95 $\pm$ 0.19 & 20.70 $\pm$ 0.03 & 19.57 $\pm$ 0.02 & 19.44 $\pm$ 0.06
 & 7 \cr
112253.50$+$005329.8 &  4.56 $\pm$ 0.02
 & 23.28 $\pm$ 0.59 & 22.73 $\pm$ 0.19 & 20.26 $\pm$ 0.03 & 19.12 $\pm$ 0.02 & 19.05 $\pm$ 0.05
 & 7 \cr
113559.94$+$002422.8 &  4.04 $\pm$ 0.02
 & 23.60 $\pm$ 0.37 & 21.35 $\pm$ 0.04 & 20.02 $\pm$ 0.02 & 19.87 $\pm$ 0.02 & 19.71 $\pm$ 0.08
 & 9 \cr
123937.18$+$674020.8*&  4.40 $\pm$ 0.02
 & 25.83 $\pm$ 0.66 & 23.71 $\pm$ 0.29 & 20.79 $\pm$ 0.04 & 20.18 $\pm$ 0.04 & 19.72 $\pm$ 0.11
 &   \cr
\noalign{\medskip\hrule}}

\clearpage

\halign{\hskip 12pt
# \hfil \tabskip=1em plus1em minus1em&
\hfil # \hfil &
\hfil # &
\hfil # &
\hfil # &
\hfil # &
\hfil # &
\hfil # \cr
\multispan8{\hfil TABLE 1. Positions and Photometry of SDSS
 High-Redshift Quasars (continued) \hfil}\cr
\noalign{\bigskip\hrule\smallskip\hrule\medskip}
\hfil Quasar (SDSSp J) \hfil&\hfil Redshift \hfil 
&\hfil $u^*$ \hfil&\hfil $g^*$ \hfil &
\hfil $r^*$ \hfil & \hfil $i^*$ \hfil & \hfil $z^*$ \hfil & \hfil Note \hfil \cr
\noalign{\medskip\hrule\bigskip}
125433.57$-$003922.7 &  4.23 $\pm$ 0.03
 & 23.50 $\pm$ 0.53 & 22.42 $\pm$ 0.10 & 20.58 $\pm$ 0.03 & 20.10 $\pm$ 0.03 & 19.99 $\pm$ 0.08
 &   \cr
125759.22$-$011130.3 &  4.10 $\pm$ 0.01
 & 24.39 $\pm$ 0.53 & 20.40 $\pm$ 0.02 & 19.00 $\pm$ 0.01 & 18.55 $\pm$ 0.01 & 18.45 $\pm$ 0.03
 &   \cr
130216.13$+$003032.1 &  4.51 $\pm$ 0.04
 & 23.70 $\pm$ 0.52 & 22.95 $\pm$ 0.18 & 20.98 $\pm$ 0.04 & 19.86 $\pm$ 0.02 & 19.69 $\pm$ 0.08
 &   \cr
131052.51$-$005533.2 &  4.16 $\pm$ 0.01
 & 23.19 $\pm$ 0.81 & 20.82 $\pm$ 0.03 & 18.86 $\pm$ 0.01 & 18.83 $\pm$ 0.01 & 18.79 $\pm$ 0.04
 & 7 \cr
132110.82$+$003821.7 &  4.67 $\pm$ 0.02
 & 23.94 $\pm$ 0.62 & 23.28 $\pm$ 0.20 & 21.48 $\pm$ 0.07 & 20.04 $\pm$ 0.03 & 20.08 $\pm$ 0.10
 & 7 \cr
\noalign{\smallskip}
132447.26$-$031358.3 &  4.02 $\pm$ 0.01
 & 23.12 $\pm$ 0.47 & 21.50 $\pm$ 0.06 & 19.98 $\pm$ 0.02 & 19.79 $\pm$ 0.03 & 19.60 $\pm$ 0.15
 &   \cr
134723.09$+$002158.9 &  4.27 $\pm$ 0.01
 & 23.59 $\pm$ 0.55 & 21.15 $\pm$ 0.03 & 19.34 $\pm$ 0.01 & 18.96 $\pm$ 0.01 & 18.85 $\pm$ 0.04
 &   \cr
135057.86$-$004355.3 &  4.40 $\pm$ 0.01
 & 24.03 $\pm$ 0.55 & 22.36 $\pm$ 0.09 & 20.29 $\pm$ 0.02 & 19.91 $\pm$ 0.02 & 19.64 $\pm$ 0.06
 &   \cr
135134.46$-$003652.2 &  4.04 $\pm$ 0.01
 & 23.16 $\pm$ 0.45 & 21.92 $\pm$ 0.06 & 20.14 $\pm$ 0.02 & 19.88 $\pm$ 0.03 & 19.75 $\pm$ 0.08
 &   \cr
135423.00$-$003906.2 &  4.42 $\pm$ 0.01
 & 24.19 $\pm$ 0.56 & 22.92 $\pm$ 0.14 & 20.64 $\pm$ 0.03 & 20.15 $\pm$ 0.03 & 19.92 $\pm$ 0.07
 &   \cr
\noalign{\smallskip}
141315.36$+$000032.4 &  4.07 $\pm$ 0.01
 & 23.68 $\pm$ 0.57 & 21.38 $\pm$ 0.05 & 19.76 $\pm$ 0.02 & 19.73 $\pm$ 0.02 & 19.75 $\pm$ 0.10
 & 7 \cr
141332.36$-$004909.7*&  4.14 $\pm$ 0.02
 & 25.35 $\pm$ 0.93 & 21.09 $\pm$ 0.04 & 19.58 $\pm$ 0.01 & 19.30 $\pm$ 0.02 & 19.09 $\pm$ 0.05
 & 7 \cr
144231.73$+$011055.3 &  4.56 $\pm$ 0.03
 & 24.00 $\pm$ 0.39 & 22.86 $\pm$ 0.13 & 20.87 $\pm$ 0.04 & 19.93 $\pm$ 0.03 & 19.62 $\pm$ 0.08
 &   \cr
144407.63$-$010152.8 &  4.51 $\pm$ 0.01
 & 24.41 $\pm$ 0.52 & 23.01 $\pm$ 0.23 & 20.32 $\pm$ 0.03 & 19.29 $\pm$ 0.02 & 19.11 $\pm$ 0.05
 &   \cr
144617.35$-$010131.2 &  4.13 $\pm$ 0.01
 & 24.69 $\pm$ 0.44 & 21.06 $\pm$ 0.04 & 19.53 $\pm$ 0.01 & 19.09 $\pm$ 0.01 & 18.89 $\pm$ 0.04
 &   \cr
\noalign{\smallskip}
152443.19$+$011358.9 &  4.10 $\pm$ 0.02
 & 23.54 $\pm$ 0.44 & 22.12 $\pm$ 0.08 & 20.17 $\pm$ 0.02 & 19.98 $\pm$ 0.03 & 19.62 $\pm$ 0.09
 &   \cr
152740.52$-$010602.7 &  4.40 $\pm$ 0.01
 & 24.15 $\pm$ 0.53 & 22.73 $\pm$ 0.15 & 20.48 $\pm$ 0.03 & 19.95 $\pm$ 0.02 & 19.63 $\pm$ 0.07
 & 7 \cr
160501.21$-$011220.6*&  4.92 $\pm$ 0.02
 & 24.50 $\pm$ 0.48 & 26.19 $\pm$ 0.38 & 22.50 $\pm$ 0.15 & 19.78 $\pm$ 0.02 & 19.87 $\pm$ 0.08
 & 7 \cr
162048.74$+$002005.7*&  4.09 $\pm$ 0.02
 & 24.30 $\pm$ 0.41 & 21.99 $\pm$ 0.06 & 19.85 $\pm$ 0.02 & 19.36 $\pm$ 0.02 & 18.95 $\pm$ 0.05
 &   \cr
170804.91$+$602202.0 &  4.35 $\pm$ 0.02
 & 23.18 $\pm$ 0.75 & 22.47 $\pm$ 0.10 & 20.24 $\pm$ 0.02 & 19.78 $\pm$ 0.03 & 19.62 $\pm$ 0.09
 &   \cr
\noalign{\smallskip}
171014.52$+$592326.5 &  4.47 $\pm$ 0.03
 & 23.03 $\pm$ 0.42 & 22.50 $\pm$ 0.11 & 20.48 $\pm$ 0.03 & 19.67 $\pm$ 0.02 & 19.36 $\pm$ 0.07
 &   \cr
171224.92$+$560625.0 &  4.20 $\pm$ 0.01
 & 23.72 $\pm$ 0.62 & 22.93 $\pm$ 0.19 & 20.54 $\pm$ 0.03 & 20.02 $\pm$ 0.04 & 20.08 $\pm$ 0.11
 &   \cr
171530.49$+$645319.3 &  3.96 $\pm$ 0.01
 & 22.64 $\pm$ 0.54 & 20.85 $\pm$ 0.04 & 19.56 $\pm$ 0.02 & 19.46 $\pm$ 0.02 & 19.49 $\pm$ 0.11
 &   \cr
171808.67$+$551511.2 &  4.60 $\pm$ 0.01
 & 24.24 $\pm$ 0.50 & 23.56 $\pm$ 0.25 & 21.15 $\pm$ 0.05 & 19.99 $\pm$ 0.03 & 19.65 $\pm$ 0.08
 &   \cr
172007.20$+$602823.9 &  4.40 $\pm$ 0.01
 & 23.24 $\pm$ 1.21 & 23.09 $\pm$ 0.21 & 20.47 $\pm$ 0.04 & 20.20 $\pm$ 0.04 & 20.06 $\pm$ 0.12
 &   \cr
\noalign{\smallskip}
173744.87$+$582829.5 &  4.94 $\pm$ 0.05
 & 24.69 $\pm$ 0.31 & 24.92 $\pm$ 0.44 & 20.93 $\pm$ 0.05 & 19.33 $\pm$ 0.03 & 18.95 $\pm$ 0.07
 &   \cr
220008.66$+$001744.8 &  4.77 $\pm$ 0.01
 & 25.20 $\pm$ 0.35 & 24.28 $\pm$ 0.38 & 20.68 $\pm$ 0.03 & 19.15 $\pm$ 0.01 & 19.31 $\pm$ 0.06
 & 1 \cr
221644.02$+$001348.3 &  4.99 $\pm$ 0.01
 & 24.02 $\pm$ 0.59 & 24.49 $\pm$ 0.46 & 21.78 $\pm$ 0.07 & 20.30 $\pm$ 0.04 & 20.33 $\pm$ 0.15
 & 1 \cr
222050.80$+$001959.1 &  4.68 $\pm$ 0.02
 & 24.71 $\pm$ 0.52 & 24.99 $\pm$ 0.57 & 21.67 $\pm$ 0.07 & 20.21 $\pm$ 0.03 & 20.17 $\pm$ 0.16
 & 1 \cr
234147.27$+$001551.9 &  3.95 $\pm$ 0.01
 & 24.16 $\pm$ 0.71 & 21.63 $\pm$ 0.06 & 19.96 $\pm$ 0.02 & 20.03 $\pm$ 0.04 & 19.73 $\pm$ 0.13
 &   \cr
\noalign{\smallskip}
235718.36$+$004350.4 &  4.35 $\pm$ 0.01
 & 24.00 $\pm$ 0.78 & 22.44 $\pm$ 0.14 & 20.07 $\pm$ 0.03 & 19.83 $\pm$ 0.04 & 19.61 $\pm$ 0.10
 & 6 \cr
\noalign{\medskip\hrule}}

\medskip\noindent
Notes: Positions are in J2000.0 coordinates.
An asterisk following an object name indicates a probable BAL quasar.
Photometry is reported in terms of asinh magnitudes; see
Lupton, Gunn, \& Szalay~(1999) for details.
In this system, zero flux
corresponds to 24.6, 25.1, 24.8, 24.4, and~22.8
in $u^*$, $g^*$,$r^*$, $i^*$, and~$z^*$, respectively.

\noindent
 1) Spectra obtained with APO 3.5-m \\
 2) Unpublished PSS Quasar; see
 {\tt http://astro.caltech.edu/$\sim$george/z4.qsos} \\
 3) PSS Quasar; Kennefick et al. (1995a) \\
 4) BRI Quasar; Finding Chart in Smith et al. (1994b) \\
 5) BRI Quasar; Finding Chart in Smith et al. (1994a) \\
 6) SDSS Quasar; Fan et al. (1999) \\
 7) SDSS Quasar; Fan et al. (2000a) \\
 8) SDSS Quasar; Fan et al. (2001) \\
 9) SDSS Quasar; Schneider et al. (2001) \\
\end{scriptsize}
\clearpage

%

\end{document}